\documentstyle[prd,aps,epsf]{revtex}

\begin{document}
\newcommand\1{$\spadesuit$}
\tighten
\draft
\twocolumn[\hsize\textwidth\columnwidth\hsize\csname 
@twocolumnfalse\endcsname

\title{Non-Vacuum Initial States \\
for Cosmological Perturbations of Quantum-Mechanical Origin}

\author{J\'er\^ome Martin${}^1$, Alain Riazuelo${}^1$ 
and Mairi Sakellariadou${}^{2,3}$}

\address{~\\
${}^1$DARC, Observatoire de Paris, \\ 
UPR 176 CNRS, 92195 Meudon Cedex, France. 
\\~\\ 
${}^2$Theory Division, CERN, CH-1211 Geneva 23, Switzerland. 
\\~\\ 
${}^3$Centre for Theoretical Physics, University of Sussex,\\
Brighton, Falmer BN1 9QH, United Kingdom.
\\~}
\date{April 1, 1999}

\maketitle

\begin{abstract}

In the context of inflation, non-vacuum initial states for
cosmological perturbations that possess a built in scale are
studied. It is demonstrated that this
assumption leads to a  falsifiable  class of models. The question of
whether they lead to conflicts with the available observations is
addressed. For this purpose, the power spectrum of the Bardeen
potential operator is calculated and compared with the CMBR anisotropies
measurements and the redshift surveys of galaxies and clusters of galaxies.
Generic predictions
of the model are: a high first acoustic peak, the presence of a bump in
the matter power spectrum and non-Gaussian statistics. The details are
controlled by the number of quanta in the non-vacuum initial
state. Comparisons with observations show that there exists a window
for the free parameters such that good agreement between the data and
the theoretical predictions is possible. However, in the case where
the initial state is a state with a fixed number of quanta, it is
shown that this number cannot be greater than a few. On the other
hand, if the initial state is a quantum superposition, then a larger
class of initial states could account for the observations,  even though
the state cannot be too different from the vacuum. 
Planned missions such as the MAP and Planck satellites and the Sloan Survey,
will demonstrate whether the new class of models proposed here
represents a viable alternative to the standard theory.

\end{abstract}

\narrowtext
\vspace{1cm}]

\def\lsim{\lower2pt\hbox{$\buildrel{<}\over{\sim}$}}
\def\Mpc{{\rm Mpc}}

\section{Introduction}
 
The observed large-scale structure in the Universe has been currently
addressed, within the framework of gravitational instability, by two
families of models: initial density perturbations can either be due to
``freezing in'' of quantum fluctuations of a scalar field (inflaton)
during an inflationary era~\cite{infl}, or they may be seeded by a
class of topological defects,  naturally formed during a 
symmetry-breaking phase transition in the early Universe~\cite{kibble}.  The
recent bulk of observational and experimental data and, in particular,
the cosmic microwave background anisotropy measurements, and the
redshift surveys of the distribution of galaxies and clusters of
galaxies, impose severe constraints on the two families of models, as
well as on the variety of possible scenarios introduced within each
family.  
\par 
The simplest topological defects models of structure
formation show conflicts with observational data. As  was first
shown in Ref.~\cite{ram}, global topological defects models predict
strongly suppressed acoustic peaks. While on large angular scales the
predicted cosmic microwave background radiation (CMBR) spectrum is in
good agreement with COBE measurements, on smaller angular scales the
topological defects models cannot reproduce the data of the Saskatoon
experiment.  One can manufacture models~\cite{ruthmairi} with
structure formation being induced by scaling seeds, which lead to an
angular power spectrum with the same characteristics (position and
amplitude of acoustic peaks), as the one predicted by standard
inflationary models. The open question is, though, whether such models
are the outcome of a realistic theory. However, the most severe
problem for topological defects models of structure formation is their
predicted~\cite{pst}, ~\cite{abr} lack of large-scale power in the matter power
spectrum, once normalized to COBE. 
 Choosing scales of $100 h^{-1}\Mpc$, which are most probably
unaffected by non-linear gravitational evolution, standard topological
defect models, once normalized to COBE, require a bias factor
($b_{100}$) on scales of $100 h^{-1}\Mpc$ of $b_{100}\approx 5$, to
reconcile the predictions for the density field fluctuations with the
observed galaxy distribution. However, the latest theoretical and
experimental studies favour a current value of $b_{100}$ close
to unity.
\par 
In what follows, we shall place ourselves within the framework of
cosmological perturbations of quantum-mechanical origin in the context
of inflationary models. The inflationary paradigm was proposed in
order to explain the shortcomings of the standard (Big Bang)
cosmological model. In addition, it offers a scenario for the
generation of the primordial density perturbations, which can lead to
the formation of the observed large-scale structure.
\par 
The theory of cosmological perturbations of quantum-mechanical origin
rests on two well-established theories. On the one hand, (linearized)
general relativity allows a calculation of the evolution and the
amplification of perturbations throughout the cosmic evolution; the
mechanism at work being parametric amplification of the fluctuations
due to the interaction of the perturbations with the
background~\cite{Grireview}. On the other hand, quantum field theory
permits to understand the origin of these perturbations. If the
quantum fields are initially, i.e. at the beginning of inflation,
placed in the vacuum state, then because of the Heisenberg principle,
fluctuations are unavoidable. Moreover, the amplitude of these
fluctuations is completely fixed.
\par 
Inflation, employing the theory of cosmological perturbations of
quantum-mechanical origin, leads to definite predictions for the
anisotropies of the CMBR, as well as for the power spectrum, which can
be tested against experimental and observational data. In particular,
simple models predicts a scale-invariant spectrum, with, provided the quantum
fields are initially placed in the vacuum, Gaussian fluctuations.
\par 
Let us briefly discuss the observational data, namely the CMBR
anisotropies measurements and the redshift
surveys of the distribution of galaxies.  
\par 
The CMBR, last scattered at the epoch
of decoupling, has to a high accuracy a black-body distribution
\cite{bb}, with a temperature $T_0=2.728\pm 0.002 ~{\rm K}$, which is
almost independent of direction.  The DMR experiment on the COBE
satellite measured a tiny variation in intensity of the CMBR, at fixed
frequency. This is equivalently expressed as a variation ${\rm \delta
}T$ in the temperature, which was measured to be 
${\rm \delta }T/T_0 \approx 10^{-5}$~\cite{cobe}.  
The 4-year COBE data are fitted by a scale-free
spectrum; the spectral index was found to be $n_{\rm S}=1.2\pm 0.3$
and the quadrupole anisotropy $Q_{\rm rms-PS}=15.3^{+3.8}_{-2.8} ~{\rm
\mu K}$~\cite{cobe}. The CMBR anisotropies spectrum is usually
parametrized in terms of the multipole moments $C_\ell$, defined as
the coefficients in the expansion of the temperature autocorrelation
function
\begin{eqnarray}
\label{C_l}
\left\langle{{\rm \delta} T\over T}({\bf e}_1)
            {{\rm \delta} T\over T}({\bf e}_2)\right\rangle
\bigg|_{({\bf e}_1\cdot {\bf e}_2=\cos\vartheta)} = 
  \nonumber \\
  {1\over 4\pi}\sum_\ell (2\ell & + & 1)\,C_\ell P_\ell(\cos\vartheta)\;~,
\end{eqnarray}
which compares points in the sky separated by an angle $\vartheta$.
The value of $C_\ell$ is determined by fluctuations on angular scales
of order $\pi/\ell$. The angular power spectrum of anisotropies
observed today is usually given by the power per logarithmic interval
in $\ell$, plotting $\ell(\ell+1)C_\ell$ versus $\ell$.
\par
On large angular scales, the main physical mechanism which contributes
to the redshift of photons propagating in a perturbed Friedmann
geometry, originates from fluctuations in the gravitational potential
on the last-scattering surface. The COBE-DMR experiment, which
measured CMBR anisotropies on such large angular scales ($\ell ~ \lsim 20$),
confirmed the predicted scale-invariant spectrum and yields
 mainly a normalization for the different
models of large-scale structure formation.
\par
On intermediate angular scales, $0.1^\circ\stackrel{<}{\sim}
\vartheta\stackrel{<}{\sim} 2^\circ$, the main contribution to the
CMBR anisotropies comes from the intrinsic inhomogeneities on the
surface of the last scattering, due to acoustic oscillations in the
coupled baryon-radiation fluid prior to decoupling.  On the same
angular scales, there is a Doppler contribution to the CMBR
anisotropies, due to the relative motions of emitter and observer. The
sum of these two contributions is denoted by the term acoustic peaks.
An analysis of recent CMBR flat-band measurements on intermediate
angular scales gives~\cite{charley} in the best-fit power spectrum a
peak $[\ell(\ell+1)C_\ell/2\pi]^{1/2}T_0= 76 ~{\rm \mu K}$ with
$\ell=260$.
\par
Among the various experiments measuring CMBR anisotropies, the
Saskatoon experiment~\cite{sask} is of particular importance since it
relates~\cite{einasto} CMBR anisotropies to the power spectrum of
matter density perturbations estimated through clustering properties
of galaxies and clusters of galaxies. More precisely, the Saskatoon
experiment measures temperature anisotropies for multipoles in the
range $\ell \approx 80 - 400$, which corresponds to the range of
wavelengths for which we have data on galaxy clusters.
\par
Analysing a large number of available data on redshifts of individual
galaxies and Abell galaxy clusters, one obtains~\cite{einasto} the
power spectrum for clusters of galaxies, over the wave-number interval
from $k\approx 0.03 ~h$~Mpc${}^{-1}$ to $k\approx 0.3 ~h~$Mpc${}^{-1}$.  On
very large scales ($k<0.03 ~h$~Mpc${}^{-1}$), the large error bars
are due to incomplete data. However, near the turn over, error bars are
small, thus both the relative position and amplitude of the turn over
are determined accurately. As discussed in, e.g. Ref.~\cite{einasto},
the power spectrum reveals the existence of a non-trivial feature at a
wave-number $k_0 =0.052\pm0.005 ~h~\Mpc^{-1}$. Assuming this peak exists
(further studies are necessary to confirm it), the amplitude
of the observed power spectrum is larger near the peak by a factor
$1.4$ \cite{einasto} with respect to the power spectrum of the
standard cold dark matter model. The existence of this peak is not
related~\cite{einasto2} to acoustic oscillations in the tight coupled
baryon--photon plasma. As stated in Ref.~\cite{einasto2}, the current
CMBR experimental data combined with observational cluster data,
favour theoretical models that have built-in a characteristic scale
in their initial spectrum.
\par
Recently, the COBE data have also been used to test the gaussianity of
the CMBR anisotropies. Three
groups~\cite{non-gaus1,non-gaus2,non-gaus3} have now reported results
showing that the fluctuations would not be Gaussian. The three groups
work with different methods. In Ref.~\cite{non-gaus1}, the estimation
of the bispectrum $B_\ell$ is used as a criterion to test
gaussianity. The dominant non-Gaussian contribution has been found
near $\ell=16$. It is clear that these results should be taken
cautiously since, for example, the issue of foreground contamination
could change the conclusions. However, the possibility of non-Gaussian
statistics in the CMBR anisotropies should be taken  into
account seriously.
\par
The Saskatoon measurements could be explained by playing with the
values of the cosmological parameters. In particular, the value of the
cosmological constant, $\Omega _{\Lambda }\approx 0.6$, recently
inferred from the SNIa measurements could account for the high position
of the first acoustic peak. But the other features (the presence of a
peak in the power spectrum; non-Gaussian fluctuations in the CMBR), if
confirmed, clearly go beyond the paradigm of cold dark matter (CDM)
and slow roll inflation. In order to explain them, different
mechanisms have been advocated. For instance, double-inflation
\cite{di} or multiple-inflation~\cite{mi} models have been used to
explain the presence of the peak in the power spectrum. Another
scenario can be offered within models where the inflaton field evolves
through a kink in its potential~\cite{kink}. To explain the 
non-Gaussianity, different mechanisms have been proposed. Of course, this
appears naturally when the perturbations are induced by topological
defects. However, even in the context of inflation, non-Gaussianity
can be present, as for example in the case of stochastic
inflation~\cite{Alejandro,stochastic}.
\par
All these solutions are in fact different possible modifications of
the power spectrum of the primordial fluctuations. In this article,
our aim is to discuss the choice of the initial quantum state in which
the quantum fields are placed. This choice is of course crucial for
the determination of the primordial power spectrum, and different
quantum states will lead to different power spectra. In the
literature, it is (almost, see Ref.~\cite{Staro}) always assumed that
the state of the perturbations is the vacuum (In a curved spacetime 
the definition of the vacuum state is not unique. A more precise 
definition of the vacuum used in this paper is given in what follows
 and coincides with the one in Ref.~\cite{rbv}):
\begin{equation}
\label{defvacuum}
|0\rangle \equiv \bigotimes _{\bf k} |0_{\bf k}\rangle .
\end{equation}
Let us examine how this choice can be justified. Since this question
is a problem of boundary conditions, it must be addressed by means of
a theory of the initial conditions for the early Universe. Such a
theory should rely on full quantum gravity, which is 
unknown at present. The only candidate at our disposal is quantum
cosmology. Generally, it predicts that the initial state is indeed the
vacuum. For example, the no-boundary choice for the wave function of
the cosmological perturbations implies that the Bardeen operator is
placed in the vacuum state, see Ref.~\cite{HHalli}. This result does
not come as a surprise since the Hartle--Hawking proposal is a
generalization of a method that gives the ground-state wave function
of a system in ordinary quantum mechanics.
\par
However, although fascinating, quantum cosmology is not yet a
well-developed branch of physics and many important questions remain
unsolved to this day. To our knowledge, there exists no proof that
quantum considerations automatically lead to a vacuum initial state
for the perturbations. Such a proof, if it exists, should rely on full
quantum gravity.
\par
On the other hand, the choice of the vacuum is also based on the
hypothesis that the initial state of the Universe should be a
``maximally symmetric state''~\cite{mssstaro}. Concretely, this means
that no scale should be privileged. This seems to be the  simplest
starting point. However, since the choice of the initial state is
supposed to appear naturally in the context of quantum gravity, it
could also be argued that such a privileged scale does exist and is equal
to the Planck scale, $l_{\rm Pl}=(\hbar G/c^3)^{1/2}
\approx 10^{-33} ~$cm. This becomes even more intriguing if one recalls 
that in order to solve the usual problems of the standard model of
cosmology, one needs $60$ $e$-folds during inflation. This means that
the Planck scale has now been stretched to a scale of at least $60$
pc. Accordingly, all the wavelengths below $60$ pc were sub-Planckian
at the moment of their generation. Of course, the structure of
space-time below the Planck scale is unknown and it may be very
different from the one we are used to. Probably, such notions as
sub-Planckian wavelengths or even scale factor are meaningless in a
regime where the gravitational quantum effects are important.
\par
The arguments presented in the previous discussion show that it is
worth studying non-vacuum initial states for cosmological
perturbations. Rather than relying on theoretical arguments, our goal
will be to allow for the possibility of non-vacuum initial states and
to establish the consequences for the observables described at the
beginning of the introduction. We study whether choices other than the
vacuum automatically lead to inconsistencies or conflicts with the
available observations or if, on the contrary, there exists a window for
the free parameters of the model, which fits the observational data.
\par
 Our choice of a non-vacuum initial state is guided by a very simple
idea: the initial state could have a built-in characteristic
scale since this seems to be the simplest way to generalize the vacuum 
state. The question now arises as for the physical origin of 
this scale. A possible answer is that the natural scale is the Planck
length stretched by the cosmological expansion. It is clear that, so as not to 
be in conflict with
observations, we would like this fundamental scale to be now
translated to the characteristic scale $l_{\rm C}\approx
200\,\Mpc=6.2\cdot 10^{26}$cm (here, as in the rest of this article we
take $h=0.5$). Since the ratio $l_{\rm C}/l_{\rm Pl}$ is given by
$l_{\rm C}/l_{\rm Pl}\approx 10^{27}e^N$, where $N$ is the number of
$e$-folds during inflation, this means that $N\approx
75$. Interestingly enough, we note that this leads to a number of
$e$-folds greater than the minimum number required, i.e. $60$. In the 
context of Linde's chaotic inflation, it is assumed that, initially, 
the inflaton potential $V(\varphi )$ is such 
that $V(\varphi _i) \approx m_{\rm Pl}^4$ where $m_{\rm Pl}$ is the 
Planck mass. If the potential is given by 
e.g. $V(\varphi )=(\lambda /4!)\varphi ^4$, this leads to an initial value of the 
scalar field greater than $4.4m_{\rm Pl}$, which is  needed to get the usual $60$ 
$e$-folds. Consequently, this  leads to   a huge number of $e$-folds, $N\approx 10^8$. It 
is clear that, with such a number, the Planck length cannot be stretched to 
$200\, \Mpc$ presently. Let us note that  these   models (with a large 
number of $e$-folds) suffer from the ``super-Planck scale 
problem''~\cite{Robert}: all the scales of cosmological interest now 
were sub-Planckian at the beginning of inflation. Since quantum field theory 
is expected to break down in this regime, the predictions of these 
models could be questionable. On the other hand, in the spirit of chaotic 
inflation itself, there exists regions of space in which the initial value of the 
field was $\varphi _i\approx 4.9m_{\rm Pl}$. This value leads to a number 
of $e$-folds equal to $75$. Therefore, the model presented in this article 
is certainly more relevant in the case where inflation does not last  for  a 
long period. In chaotic inflation, the probability of having a long period 
of inflation is greater than the probability to get a small 
number of $e$-folds. Thus, our model does not fit very well within the  
chaotic 
inflation approach.
\par 
 It should also be mentioned that it has been shown in 
Ref.~\cite{Robertinistate} that a large class of initial states approaches 
the Bunch Davis vacuum in the de Sitter spacetime. However, this class of initial 
states has a Gaussian wave functional and therefore the argument that the choice 
of a non-vacuum initial state would involve exponential fine tunning does not apply 
to the case considered here.
\par
Recently, a model with a small number of $e$-folds has been constructed 
in Ref.~\cite{Sarkar}. This kind of models naturally arises in the context 
of supersymetric (SUSY and SUGRA) inflation. They are particulary well 
suited to the model put forward in this article. They consist in 
multiple bursts of inflation which in total last for $\approx 75$ 
expansion times. In addition, the last stage of inflation is preceded 
by other inflationary epochs. The ``initial state'' of this last 
epoch is the result of the evolution of the ``true initial state'' through 
the multiple preceding bursts of inflation. Clearly, there is no 
reason for assuming   this ``effective initial state'' to  be the vacuum. Let 
us emphasize that this argument holds for every model with many stages 
of inflation since, in this case, the origin of the characteristic 
scale could no longer be the Planck length stretched to $l_{\rm C}$ but 
could correspond to the time where one of the fields starts rolling down.
\par
Our model possesses a privileged scale and therefore  belongs to the
class of models already envisaged in Ref.~\cite{bsi}. However, we
would like to emphasize that the origin of this scale is physically
completely different and we will point out that there exist
observables, which, in principle, allow us to distinguish between the
different models. A last comment on the fine tuning issue is in 
order here. It is true that the position of the characteristic scale must 
be chosen carefully. Otherwise the model would simply 
be in contradiction with the available data. We would like to 
emphasize that this is not a feature of our model only but in fact 
of all the BSI  (broken scale invariant)  models~\cite{bsi}. In this 
respect our model is similar to the other BSI models.
\par
This paper is organized as follows: In Section II we discuss non-vacuum 
initial state for the cosmological perturbations. We first
briefly describe the theory of perturbations of quantum-mechanical
origin. We then describe the non-vacuum initial states considered in
this article. We finally calculate the power spectra of the Bardeen
potential for these states  and show that it possesses either 
a step or a bump. In Section III we examine the
observational consequences of the calculated power spectra; we compare
our theoretical predictions with current experimental and
observational data, which will fix the parameters of our model. We end
with the conclusions given in Section IV.

\section{Non-vacuum initial state for the perturbations}

\subsection{Perturbations of quantum-mechanical origin}

The background model is described by a 
Friedmann--Lema\^{\i}tre--Robertson--Walker (FLRW) metric whose space-like sections are flat
($c=1$): ${\rm d}s^2=a^2(\eta )(-{\rm d}\eta ^2 +{\rm d}{\bf
x}^2)$. We assume that inflation is driven by a single scalar field,
$\varphi _0(\eta )$. It is convenient to define the background
quantities ${\cal H}(\eta )$ and $\gamma (\eta )$ by:
\begin{equation}
\label{defH,gamma}
{\cal H}\equiv a'/a, \quad  \gamma (\eta )
\equiv 1-\frac{{\cal H}'}{{\cal H}^2},
\end{equation}
where the primes denote the derivatives with respect to conformal time. In the
case of the de Sitter space-time, $\gamma $ vanishes.
\par
In the synchronous gauge, without loss of generality, the line element
for the FLRW background plus scalar perturbations can be written
as~\cite{Griden}:
\begin{eqnarray}
\label{metricsg}
{\rm d}s^2 &=& a^2(\eta )\biggl\{-{\rm d}\eta ^2+\biggl[\delta _{ij}
+\frac{1}{(2\pi )^{3/2}} 
\int {\rm d}{\bf k}\biggl(h(\eta ,{\bf k})\delta _{ij}\nonumber \\
& &-\frac{h_l(\eta ,{\bf k})}{k^2}k_ik_j\biggr) 
e^{i{\bf k}\cdot{\bf x}}\biggr]{\rm d}x^i{\rm d}x^j\biggr\},
\end{eqnarray}
where the functions $h, h_l$ represent the scalar perturbations of the
gravitational field and the longitudinal--longitudinal perturbation,
respectively. In the same manner, the perturbations of the scalar
field are Fourier decomposed according to:
\begin{equation}
\label{defdeltasf}
{\rm \delta }\varphi (\eta ,{\bf x})=\frac{1}{(2\pi )^{3/2}}
\int {\rm d}{\bf k}\varphi _1(\eta ,{\bf k})e^{i{\bf k}\cdot {\bf x}}.
\end{equation}
The perturbed Einstein equations couple the scalar sector, $h$ and
$h_l$, to the perturbed scalar field $\varphi _1$: see
Ref.~\cite{Griden}. It can be shown that the residual gauge invariant
quantity $\mu (\eta ,{\bf k})$
\cite{Griden} defined by:
\begin{equation}
\label{defmu}
\mu \equiv \frac{a}{{\cal H}\sqrt{\gamma }}(h'+{\cal H}\gamma h),
\end{equation}
can be used to express all other relevant quantities. The quantity $\mu
(\eta ,{\bf k})$ is related to the gauge-invariant Bardeen potential
through the equation:
\begin{equation}
\label{RPhimu}
\Phi ^{({\rm SG})}=\frac{{\cal H}\gamma }{2k^2}
\biggl(\frac{\mu }{a\sqrt{\gamma }}\biggr)', 
\end{equation}
where ``{\rm SG}'' means calculated in the synchronous gauge, see
Ref.~\cite{MS1}. The quantity $\mu (\eta ,{\bf k})$ is not defined in
the de Sitter case. This case must be treated separately and for it we
have $\Phi ^{({\rm SG})}=0$: there are no density perturbations at all
because the fluctuations of the scalar field are not coupled to the
perturbations of the metric when the equation of state is
$p=-\rho$. The perturbed Einstein equations imply that the equation of
motion for $\mu (\eta ,{\bf k})$ is given by~\cite{Griden}:
\begin{equation}
\label{paraeq}
\mu ''+\biggl[k^2-\frac{(a\sqrt{\gamma })''}{(a\sqrt{\gamma })}\biggr]\mu=0.
\end{equation}
The above is the characteristic equation of a parametric
oscillator whose time-dependent frequency depends on the scale factor
and its derivative (up to $a^{(4)}$).
\par
In this article, we assume that the perturbations are of
quantum-mechanical origin. This hypothesis fixes completely the
normalization of ${\rm \delta }\varphi (\eta ,{\bf x})$ and of the
scalar perturbations. The normalization is fixed in the high-frequency
regime. In this regime, the perturbed field can be considered as a
free field in the curved FLRW background space-time. It is therefore
necessary to study the quantization of such a free field denoted in
the following by $\varphi (\eta ,{\bf x})$. We first address this
question and we then make the link between $\varphi (\eta ,{\bf x})$
and ${\rm \delta }\varphi (\eta ,{\bf x})$. We choose to define the
Fourier component of $\varphi (\eta ,{\bf x})$ according to:
\begin{equation}
\label{defchi}
\varphi (\eta ,{\bf x})=\frac{1}{a(\eta )}\frac{1}{(2\pi )^{3/2}}
\int {\rm d}{\bf k}\chi (\eta ,{\bf k})e^{i{\bf k}\cdot {\bf x}},
\end{equation}
where we have renormalized the time-dependent amplitude with the scale
factor.  The Fourier component satisfies $\chi (\eta ,-{\bf k})=\chi
^*(\eta ,{\bf k})$, because the field is real. The action, given by:
\begin{eqnarray}
\label{action1}
S &=&\int {\rm d}^4x {\cal L} \\
\label{action2}
&=&\frac{1}{2}\int {\rm d}^4x 
a^2\biggl[\left(\varphi ' (\eta ,{\bf x})\right)^2
  -\sum _i \left({\rm \partial}_i \varphi (\eta ,{\bf x})\right)^2\biggr],
\end{eqnarray}
can also be written in terms of the Fourier component $\chi (\eta
,{\bf k})$. The result reads:
\begin{eqnarray}
\label{actionfourier}
S &=& \int {\rm d}\eta \int _{R^{3+}}{\rm d}{\bf k} \bar{{\cal L}} \\
\label{actionfourier2}
&=& \int {\rm d}\eta \int _{R^{3+}}{\rm d}{\bf k}  
\biggl\{|\chi '(\eta ,{\bf k})|^2 +
\biggl(\frac{a'^2}{a^2}-k^2\biggr)|\chi (\eta ,{\bf k})|^2
\nonumber \\
& &-\frac{a'}{a}
\biggl[\chi '(\eta ,{\bf k})\chi ^*(\eta ,{\bf k}) 
+\chi '^{*}(\eta ,{\bf k})\chi (\eta ,{\bf k})\biggr]\biggr\}.
\end{eqnarray} 
The variation of the action leads to the equation of motion for the 
Fourier component: $\chi ''+\chi [k^2-a''/a]=0$. Again, we find a
parametric oscillator-type equation. Of course, if there is no
expansion, or if the Universe is in the radiation-dominated era, 
it reduces to the usual equation of motion of an harmonic oscillator.
\par
We now turn to the Hamiltonian formalism. The first step is the
calculation of the momentum conjugate to $\varphi (\eta ,{\bf x})$
defined by:
\begin{equation}
\label{defpi}
\pi (\eta, {\bf x})\equiv \frac{{\rm \partial}{\cal L}}
{{\rm \partial}(\varphi '(\eta ,{\bf x}))}=a^2\varphi '(\eta ,{\bf x}).
\end{equation}
$\pi (\eta ,{\bf x})$ can be expressed in terms of the momentum
conjugate to $\chi (\eta ,{\bf k})$,
\begin{equation}
\label{defp}
p(\eta, {\bf k})\equiv\frac{{\rm \partial}{\bar{\cal L}}}{{\rm \partial}(
\chi '^*(\eta ,{\bf k}))}
=\chi '(\eta ,{\bf k})-\frac{a'}{a}\chi (\eta ,{\bf k}),
\end{equation}
through the relation:
\begin{equation}
\label{Rpip}
\pi (\eta, {\bf x})=\frac{a(\eta )}{(2\pi)^{3/2}}\int {\rm d}{\bf k}
p(\eta, {\bf k})e^{i{\bf k}\cdot {\bf x}}.
\end{equation} 
As a preparation to the quantization, the normal variable $\alpha
(\eta ,{\bf k})$ is introduced. Its definition is given by:
\begin{equation}
\label{alpha}
\alpha (\eta ,{\bf k})\equiv 
  N(k)\biggl[\chi (\eta ,{\bf k})+\frac{i}{k}p(\eta ,{\bf k})\biggr],
\end{equation}
where $N(k)$ is, at the classical level, a free factor.
\par
We are now in a position where quantization can be carried out. The
field $\varphi (\eta ,{\bf x})$ and its conjugate momentum $\pi (\eta
,{\bf x})$ become operators that satisfy the commutation relation:
\begin{equation}
\label{commutator}
[\hat{\varphi }(\eta ,{\bf x}),\hat{\pi }(\eta, {\bf x}')]
=i\hbar\delta ({\bf x}-{\bf x}').
\end{equation}
The normal variable $\alpha (\eta ,{\bf k})$ becomes a dimensionless
operator $c_{\bf k}(\eta )$ such that, at any time, $[c_{\bf k}(\eta
), c_{{\bf k}'}^{\dagger }(\eta )] =\delta_{{\bf k}{\bf k}'}$. With
the help of Eqs.~(\ref{defchi}) and (\ref{Rpip}) and of the definition
of the normal variable, Eq.~(\ref{alpha}), the field operator and the
conjugate momentum operator can now be expressed in terms of the
annihilation and creation operators $c_{\bf k}(\eta )$ and $c_{{\bf
k}}^{\dagger }(\eta )$. The normalization factor $N(k)$ is fixed by
the following argument: the energy of the scalar field is given by
$E=\int {\rm d}^3x \sqrt{-g}\rho $ where $\rho =-T^0{}_0$ is the
time--time component of the stress--energy tensor. Requiring that $E$
takes the following (usual) suggestive form in the high-frequency
regime,
\begin{equation}
\label{Energy}
E=\int {\rm d}{\bf k}\frac{\hbar \omega }{2}[
c_{{\bf k}}c_{{\bf k}}^{\dag }+ c_{{\bf k}}^{\dag }c_{{\bf k}}],
\end{equation}
leads to
\begin{equation}
\label{N}
N(k)=\sqrt{\frac{k}{2\hbar }}.
\end{equation}
Therefore, the scalar field operator can be written as:
\begin{eqnarray}
\label{fieldoperator}
\hat{\varphi } (\eta ,{\bf x}) & =& \frac{\sqrt{\hbar }}{a(\eta )}
\frac{1}{(2\pi )^{3/2}} \nonumber \\
 & \times & \int \frac{{\rm d}{\bf k}}{\sqrt{2k}}
\biggr[c_{\bf k}(\eta )e^{i{\bf k}\cdot {\bf x}}
+c_{\bf k}^{\dag}(\eta )e^{-i{\bf k}\cdot {\bf x}}\biggl].
\end{eqnarray}
This equation no longer contains arbitrary (or unfixed) factors. The
spirit of this argument is comparable to that of the method
employed in Ref.~\cite{Griden}.
\par
The Hamiltonian can be deduced from the action in a straightforward
manner and reads:
\begin{equation}
\label{Hamiltonian}
H=\frac{\hbar }{2}\int {\rm d}{\bf k}\biggl[k
(c_{\bf k}c_{\bf k}^{\dag }+c_{-{\bf k}}^{\dag }c_{-{\bf k}})
-i\frac{a'}{a}(c_{\bf k}c_{-{\bf k}}
-c_{\bf k}^{\dag }c_{-{\bf k}}^{\dag })\biggr].
\end{equation}
In the above expression, the first term represents the Hamiltonian of
a set of harmonic oscillators, whereas the second term can be viewed
as an interaction term between the perturbations and the
background. This term is present only in a dynamical  FLRW Universe since
the coupling function is proportional to the first time derivative of
the scale factor. The time evolution of the field operator is given by
the time evolution of the creation and annihilation operators. It can
be calculated using the Heisenberg equation:
\begin{equation}
\label{Heisenberg}
i\hbar \frac{{\rm d}}{{\rm d}\eta }\hat{\varphi }(\eta ,{\bf x})
=[\hat{\varphi }(\eta ,{\bf x}), H].
\end{equation}
This equation can be solved by a Bogoliubov transformation, 
expressed as: 
\begin{eqnarray}
\label{Bog1}
c_{\bf k}(\eta ) &=& u_k(\eta )c_{\bf k}(\eta _0)+v_k(\eta )
c_{-{\bf k}}^{\dag }(\eta _0), \\
\label{Bog2}
c_{\bf k}^{\dag }(\eta ) &=& u_k^*(\eta )c_{\bf k}^{\dag }(\eta _0)
+v_k^*(\eta )c_{-{\bf k}}(\eta _0),
\end{eqnarray}
where the functions $u_k(\eta )$ and $v_k(\eta )$ only depend on the
norm of the vector ${\bf k}$. These functions are such that
$|u_k|^2-|v_k|^2=1$, so that the commutation relation between the
creation and annihilation operators is preserved in time. The time
$\eta _0$ must be thought of as the time where the initial conditions are
fixed. Whatever these last ones are, we have $u_k(\eta _0)=1$ and
$v_k(\eta _0)=0$. The differential equations that allow the
determination of $u_k(\eta )$ and $v_k(\eta )$ are:
\begin{equation}
\label{eqsuv}
iu_k'=ku_k+i\frac{a'}{a}v_k^*, \quad iv_k'=kv_k+i\frac{a'}{a}u_k^*.
\end{equation}
If we introduce the Bogoliubov transformation given by
Eqs.~(\ref{Bog1}) and(\ref{Bog2}) in the expression of the field
operator, Eq.~(\ref{fieldoperator}), we obtain:
\begin{eqnarray}
\label{foperatoruv}
\hat{\varphi }(\eta ,{\bf x}) &=& \frac{\sqrt{\hbar }}{a(\eta )}
\frac{1}{(2\pi )^{3/2}}
\int \frac{{\rm d}{\bf k}}{\sqrt{2k}}
\biggl[c_{\bf k}(\eta _0)(u_k+v_k^*)(\eta )e^{i{\bf k}\cdot {\bf x}} 
\nonumber \\
& &+c_{\bf k}^{\dag}(\eta _0)(u_k^*+v_k)(\eta )e^{-i{\bf k}\cdot 
{\bf x}}\biggr].
\end{eqnarray}
>From Eq.~(\ref{eqsuv}), it is easy to see that the function
$(u_k+v_k^*)(\eta )$ satisfies the same equation as $\chi (\eta ,{\bf
k})$. In the high-frequency regime, the term $a''/a$ becomes
negligible and we have $\lim _{k\rightarrow +\infty }(u_k+v_k^*)(\eta
)=e^{-ik(\eta -\eta _0)}$. This means that, in this regime, the
operator $\hat{\chi }(\eta ,{\bf k})$ is given by:
\begin{eqnarray}
\label{hfchi}
\lim _{k\rightarrow +\infty}\hat{\chi } (\eta ,{\bf k})
  = \qquad\qquad\qquad\qquad\qquad\qquad\qquad & & \nonumber \\
\qquad \sqrt{\hbar}
       \biggl[c_{\bf k}(\eta _0)
                \frac{e^{-ik(\eta -\eta _0)}}{\sqrt{2k}}
             +c_{-{\bf k}}^{\dagger}(\eta _0)
                \frac{e^{ik(\eta -\eta _0)}}{\sqrt{2k}}\biggr].
\end{eqnarray}
We can now come back to our initial problem, which consists in finding
the correct normalization of the perturbed scalar field and of the
scalar perturbations. We can identify the Fourier component operator
of the perturbed field, $\hat{\varphi }_1(\eta ,{\bf k})$, with
$\hat{\chi }(\eta ,{\bf k})/a(\eta )$, both operators being considered
in the high-frequency regime. Let us emphasize again that this
identification is valid only in this regime. Otherwise, the field
$\hat{\varphi }_1(\eta ,{\bf k})$ does not behave as the free field
$\hat{\varphi }(\eta ,{\bf x})$ and the time dependence of the modes
is no longer given by the function $(u_k+v_k^*)(\eta )$. The
normalization of the perturbed scalar field fixes automatically the
normalization of the scalar perturbations of the metric since they are
linked through Einstein's equations. We only need this link in the
high-frequency regime. It can be expressed as (see
Refs.~\cite{Griden,MS1}):
\begin{equation}
\label{linkmufield}
\lim _{k\rightarrow +\infty }\hat{\mu }(\eta ,{\bf k})=
-\sqrt{2\kappa }a(\eta )\lim _{k\rightarrow +\infty }\hat {\varphi }_1
(\eta ,{\bf k}),
\end{equation}
where $\kappa \equiv 8\pi G$. From the last expression and
Eq.~(\ref{hfchi}), we immediately deduce that:
\begin{eqnarray}
\label{hfmu}
\lim _{k\rightarrow +\infty }\hat {\mu }(\eta ,{\bf k})
  = \quad\qquad\qquad\qquad\qquad\qquad\qquad\qquad & & \nonumber \\
-4\sqrt{\pi }l_{\rm Pl}\biggl[c_{\bf k}(\eta _0)
\frac{e^{-ik(\eta -\eta _0)}}{\sqrt{2k}} +c_{-{\bf k}}^{\dagger }(\eta _0)
\frac{e^{ik(\eta -\eta _0)}}{\sqrt{2k}}\biggr],
\end{eqnarray}
where $l_{\rm Pl}=(G \hbar)^{1/2}$ is the Planck length. As announced,
the normalization of the scalar perturbations is now completely
determined. In general, the operator $\hat{\mu }(\eta ,{\bf k})$ will
be given by:
\begin{equation}
\hat{\mu }(\eta ,{\bf k})=-4
\sqrt{\pi }l_{\rm Pl}[c_{\bf k}(\eta _0)\xi _k(\eta )+
c_{-{\bf k}}^{\dag }(\eta _0) \xi _k^*(\eta )]~.
\end{equation}
 The function $\xi _k(\eta )$ is the solution of Eq.~(\ref{paraeq})
 such that $\lim _{k\rightarrow +\infty } \xi _k =e^{-ik(\eta -\eta
 _0)}/\sqrt{2k}$. If we introduce the function $f_k(\eta )$ defined by
\begin{equation}
f_k(\eta )\equiv 
-4\sqrt{\pi }[({\cal H}\gamma )/(2k^2)][\xi _k/(a\sqrt{\gamma })']~,
\end{equation}
then the dimensionless Bardeen operator $\hat{\Phi }(\eta ,{\bf x})$
can be written as:
\begin{eqnarray}
\label{operatorphi}
\hat{\Phi }(\eta ,{\bf x}) & = & \frac{l_{\rm Pl}}{(2\pi )^{3/2}} \\ 
 & \times & \int {\rm d}{\bf k}
\biggl[c_{\bf k}(\eta _0)f_k(\eta )e^{i{\bf k}\cdot {\bf x}}
+c_{\bf k}^{\dag }(\eta _0)f_k^*(\eta )e^{-i{\bf k}\cdot {\bf x}}\biggr].
\nonumber
\end{eqnarray}
This equation is the main equation of this section and will be used in
the following.
\par
To conclude this part, it is interesting to consider the previous
equations in the case of power law inflation, i.e. when the scale
factor is given by:
\begin{equation}
\label{defpl}
a(\eta )=l_0|\eta |^{1+\beta }.
\end{equation}
To have inflation, $\beta $ must be such that $\beta +2<0$ (the case
$-2<\beta <-1$ is not considered here because it cannot be realized
with a single scalar field); $\beta =-2$ corresponds to the de Sitter
universe; $l_0$ has dimension of length and, in the de Sitter case, it
is equal to the Hubble radius $l_H\equiv a^2/a'$.  Moreover, in a de
Sitter universe, the function $\gamma (\eta )$ turns out to be zero.
 We would like to notice
that the assumption of power-law inflation is not as restrictive as it
seems. Indeed, the widely used slow-roll approximation boils down to
power-law inflation, with an effective $\beta $ depending on the slow-roll 
parameters, see Refs.~\cite{Lea}. In the case of power law
inflation, analytical exact solutions for the equations of motion of
the perturbations can be found. With the scale factor of
Eq.~(\ref{defpl}), Eq.~(\ref{paraeq}) can be solved in terms of Bessel
functions. Then, the exact solution for the function $\xi _k(\eta )$
is:
\begin{equation}
\xi _k(\eta )=-i(\pi /2)^{1/2}e^{i(k\eta _0-\pi \beta /2)}
(k\eta )^{1/2}H^{(2)}_{\beta +1/2}(k\eta )/\sqrt{2k}~, 
\end{equation}
where $H^{(2)}_{\beta +1/2}$ is the second-kind Hankel function of
order $\beta +1/2$. It is straightforward to deduce the corresponding
equation for $f_k(\eta )$:
\begin{equation}
\label{f_kpl}
f_k(\eta ) =  
 -\pi \sqrt{2}\frac{{\cal H}\sqrt{\gamma }}{k^2}
 e^{i(k\eta_0-\frac{\pi \beta }{2})} 
 \frac{ik}{a}\frac{\sqrt{k\eta}}{\sqrt{2k}}
 H^{(2)}_{\beta +3/2}(k\eta ).
\end{equation}
The previous expression for $f_k(\eta)$ guarantees that the field
$\hat{\Phi }(\eta ,{\bf x})$ possesses the correct behaviour in the
high-frequency regime. Roughly speaking, $$\lim _{k\rightarrow +\infty
}f_k(\eta)\sim e^{-ik\eta }/\sqrt{2k},$$ with the correct
amplitude. This should be the case of any field, regardless of the initial
conditions one may choose.

\subsection{Quantum states}

The formulation of quantum field theory used in the previous
subsection was written in the Heisenberg picture. So far, we have only
calculated the time dependence of the Bardeen operator. In order to
describe completely the system, one must in addition specify in which
quantum state the field $\hat{\Phi }(\eta ,{\bf x})$ is placed. As we
already mentioned, it is usually found in the literature  that the
initial state of the perturbations is taken to be the vacuum. Here we
address the hypothesis that the perturbations are initially in a non-vacuum 
state.  Our choice of non-vacuum states is guided by the idea
that one must introduce a scale in the theory. We denote the
corresponding wave number by $k_0$. We examine three different non-vacuum 
states.
\par
Let ${\cal D}$ be a domain in the momentum space characterized by the
numbers $k_0$ and $\sigma$, such that $k_0$ is the privileged scale
and $\sigma $ the dispersion around it. Concretely, we take ${\cal D}$
as the space between the spheres of radius $k_0-\sigma $ and
$k_0+\sigma $, i.e. a shell of width $2\sigma $ in $k$-space. The
domain ${\cal D}$ is invariant under rotations and therefore is
compatible with the assumption of isotropy of the Universe. Our first
state is given by:
\begin{eqnarray}
\label{defpsi1}
|\Psi _1(k_0,\sigma ,n)\rangle & \equiv &
\prod _{{\bf k} \in {\cal D}(k_0,\sigma )}
\frac{(c_{\bf k}^{\dagger })^n}{\sqrt{n!}} |0_{\bf k}\rangle 
\bigotimes _{{\bf p}\not\in {\cal D}(k_0,\sigma )}|0 _{\bf p}
\rangle , \\
&=& \bigotimes _{{\bf k} \in {\cal D}(k_0,\sigma )}|n_{{\bf k}}\rangle 
\bigotimes _{{\bf p} \not\in {\cal D}(k_0,\sigma )}|0_{\bf p}\rangle .
\end{eqnarray} 
The state $|n_{{\bf k}}\rangle $ is an $n$-particle state satisfying,
at $\eta =\eta _0$: $c_{{\bf k}}|n_{{\bf k}}\rangle
=\sqrt{n}|(n-1)_{{\bf k}}\rangle $ and $c_{{\bf k}}^{\dag}|n_{{\bf
k}}\rangle =\sqrt{n+1}|(n+1)_{{\bf k}}\rangle $.
\par
More complicated states can be constructed by considering quantum
superpositions of $|\Psi _1(k_0,\sigma ,n)\rangle $. We will consider
the following state:
\begin{equation}
\label{defpsi2}
|\Psi _2(k_0,n)\rangle \equiv \int {\rm d}\sigma g(\sigma )
|\Psi _1(k_0,\sigma ,n)\rangle ,
\end{equation}
where, a priori, $g(\sigma )$ is an arbitrary function of $\sigma
$. It is clear from the definition of the state $|\Psi _1\rangle $
that the transition between the empty and the filled modes is
sharp. Physically, this is probably not very realistic. The function
$g(\sigma )$ will be chosen in order to ``smooth out'' the quantum
state $|\Psi _1\rangle $. Also it should be clear that the writing of
$|\Psi _2\rangle $ in Eq.~(\ref{defpsi2}) is symbolic. An accurate
definition of this state requires to take into account subtleties,
which will be considered when we calculate the spectrum in the next
section.
\par
Finally a third state can be defined according to:
\begin{equation}
\label{defpsi3}
|\Psi _3(k_0)\rangle \equiv \sum _{n=0}^{\infty }h(n)|\Psi
_2(k_0,n)\rangle.
\end{equation}
The function $h(n)$ is arbitrary. As demonstrated below, this state
will allow us to work with an effective number of quanta, which will no
longer be an integer. This state seems to be the most natural
rotational-invariant, smooth, quantum state that privileges a
scale. Typically, the quantum state given in Eq.~(\ref{defpsi3})
depends on $k_0$ and on the free parameters characterizing the
functions $g(\sigma )$ and $h(n)$. Their values will be determined
later by confronting our theoretical predictions versus experimental
and observational data.
\par
Our aim is now to calculate the power spectra of the Bardeen potential
operator in the three states $|\Psi_i\rangle $, $i=1,2,3$.

\subsection{Power spectra}

>From now on, for convenience, we consider that the system is in a box
whose edges have length $L$. As a consequence, the wave vector
possesses discrete components given by $k_i=[(2\pi )/L]m_i$, where
$m_i$ is an integer. The Bardeen operator can be written as:
\begin{eqnarray}
\label{discretephi}
\hat{\Phi }(\eta ,{\bf x})=\frac{l_{\rm Pl}}{L^{3/2}}
\sum _{\bf k}& & [c_{\bf k}(\eta _0)f_k(\eta )e^{i{\bf k}\cdot {\bf x}}
\nonumber \\
& &
+c_{\bf k}^{\dag }(\eta _0)f_k^*(\eta )e^{-i{\bf k}\cdot {\bf x}}].
\end{eqnarray}
We pass from the discrete representation to the continuous one by
sending $L$ to infinity and by applying the rule $1/(2\pi )^{3}\int
{\rm d}{\bf k}\rightarrow 1/L^3\sum _{\bf k}$. It is clear that the
final result does not depend on $L$.
\par
The power spectrum of $\hat{\Phi }(\eta ,{\bf x})$ in the state $|\Psi
\rangle$, denoted by $P_{\Phi }(k;|\Psi \rangle )$, is defined
through the calculation of the two-point correlation function
$K_2(r;|\Psi \rangle )$. In the continuous limit,
\begin{eqnarray}
\label{defK}
K_2(r;|\Psi \rangle )& \equiv &{\langle \Psi |\hat{\Phi }(\eta ,{\bf x})
\hat{\Phi }(\eta ,{\bf x}+{\bf r})|\Psi \rangle} \over 
{\langle \Psi |\Psi \rangle}
\nonumber \\ 
&=& \int _0 ^{\infty } 
\frac{{\rm d}k}{k}\frac{\sin kr}{kr}k^3P_{\Phi }(k;|\Psi \rangle).
\end{eqnarray}
In this definition, we have taken into account the fact that the state
$|\Psi \rangle $ is not automatically normalized to 1. The power
spectrum is a time-dependent function but in the long-wavelength limit
this dependence disappears. In order to perform the computation of the
correlation function for the state $|\Psi _1\rangle $, one needs the
following quantities:
\begin{eqnarray}
\label{ccpsi21}
& &\langle \Psi _1(k_0,\sigma ,n)|c_{\bf p}c_{\bf q}|
\Psi _1(k_0,\sigma ,n)\rangle 
\nonumber \\
& & =\langle \Psi _1(k_0,\sigma ,n)|c_{\bf p}^{\dag}
c_{\bf q}^{\dag}|\Psi _1(k_0,\sigma ,n)\rangle =0, \\
\label{ccpsi22}
& &\langle \Psi _1(k_0,\sigma ,n)|c_{\bf p}
c_{\bf q}^{\dag}|\Psi _1(k_0,\sigma ,n)
\rangle 
=n{\rm \delta }({\bf q}\in {\cal D}){\rm \delta }_{{\bf p}{\bf q}}
+{\rm \delta }_{{\bf p}{\bf q}}, \nonumber \\
& & \\
\label{ccpsi23}
& &\langle \Psi _1(k_0,\sigma ,n)|c_{\bf p}^{\dag }
c_{\bf q}|\Psi _1(k_0,\sigma ,n)
\rangle =n{\rm \delta }({\bf q}\in {\cal D}){\rm \delta }_{{\bf p}{\bf q}}.
\end{eqnarray} 
In these formulas, ${\rm \delta }({\bf q}\in {\cal D})$ is a function
that is equal to $1$ if ${\bf q}\in {\cal D}$ and $0$ otherwise.
\par
As a warm up, we calculate the power spectrum for the state $|\Psi
_1\rangle $ with $n=0$, i.e. for the vacuum. Using the previous
equations in the definition of the correlation function,
Eq.~(\ref{defK}), one finds:
\begin{equation}
\label{vac}
K_2(r;|0\rangle ) 
=\frac{l_{Pl}^2}{L^3}\sum _{\bf k}|f_k|^2e^{-i{\bf k}\cdot {\bf r}},
\end{equation}
which in the continuous limit $L\rightarrow +\infty$ goes to
\begin{equation}
\label{vaccont} 
\frac{l_{Pl}^2}{(2\pi )^3}\int d^3\vec{k}|f_k|^2 e^{-i{\bf k}\cdot{\bf r}}.
\end{equation}
After having performed the angular integrations, we recover the power
spectrum $P_{\Phi }(k;|0 \rangle )$:
\begin{equation}
\label{Pvac}
k^3P_{\Phi }(k;|0 \rangle )=\frac{l_{\rm Pl}^2}{2\pi ^2}k^3|f_k|^2.
\end{equation}
Let us turn to the calculation of $K_2(r;|\Psi _1\rangle $. It can be
expressed as:
\begin{equation}
\label{Kpsi1}
K_2(r;|\Psi _1\rangle ) 
=\frac{l_{Pl}^2}{L^3}\sum _{\bf k }\biggl( |f_k|^2e^{-i{\bf k}\cdot {\bf r}}
[1+2n{\rm \delta }({\bf k}\in {\cal D})]\biggr).
\end{equation}
>From this equation and from the definition of the function ${\rm
\delta }({\bf k}\in {\cal D})$, we deduce the expression of the
power spectrum:
\begin{eqnarray}
\label{Ppsi1}
k^3P_{\Phi }(k;|\Psi _1 \rangle ) &=& \frac{l_{\rm Pl}^2}{2\pi ^2}
k^3|f_k|^2\biggl\{1 +2n\nonumber \\
&\times & [{\rm H}(k-k_0+\sigma )-{\rm H}(k-k_0-\sigma )]\biggr\},
\end{eqnarray}
where $\rm H$ is a Heaviside function. We see that, in addition to the
usual vacuum spectrum, there is a new contribution located around the
wave number $k_0$. This new contribution vanishes if $n=0$, as
expected.
\par
This spectrum is not continuous. As already mentioned, this is not
physically very realistic. It has for origin the very crude definition
of the state $|\Psi _1 \rangle $. We thus turn to the case where the
quantum state is given by $|\Psi _2 \rangle $. This refinement will
allow us to obtain a smooth and physical spectrum.
\par
Since the system is placed in a box, the state $|\Psi _2\rangle $ can
be defined by a discrete sum according to:
\begin{equation}
\label{defpsi2dis}
|\Psi _2(k_0,n)\rangle \equiv 
 \sum _{i=0}^{N} g_i|\Psi _1(k_0,\sigma _i,n)\rangle ,
\end{equation}
where $g_i$ and $\sigma _i$, $i=0, \dots ,N$, are just series of
numbers. We choose the $\sigma _i$'s such that:
\begin{equation}
\label{orthocondition}
\langle \Psi _1(k_0,\sigma _i,n)|\Psi _1(k_0,\sigma _j,n)\rangle
 =\delta _{ij}.
\end{equation}
This is satisfied if the number of modes in the domains ${\cal
D}(k_0,\sigma _i)$ and ${\cal D}(k_0,\sigma _j)$, ${\cal N}({\cal
D}_i)$ and ${\cal N}({\cal D}_j)$ respectively, are such that: ${\cal
N}({\cal D}_i)-{\cal N}({\cal D}_j)\ge 1$. This condition boils down
to $\sigma _i -\sigma _j \ge \pi ^2/ [L^3(k_0^2+\sigma _i^2)]$, where
we have assumed that $\sigma _i- \sigma _j\ll 1$. Therefore, we can
always find a value of $L$ such that the condition be fulfilled. Then,
the calculation of $\langle \Psi _2|\hat{\Phi }(\eta ,{\bf
x})\hat{\Phi }(\eta ,{\bf x}+{\bf r})|\Psi _2\rangle $ can be
performed. The result reads:
\begin{eqnarray}
\label{Kpsi2}
& &\langle \Psi _2|\hat{\Phi }(\eta ,{\bf x})
\hat{\Phi }(\eta ,{\bf x}+{\bf r})|\Psi _2\rangle 
=\frac{l_{Pl}^2}{L^3}\sum _{\bf k }
|f_k|^2e^{-i{\bf k}\cdot {\bf r}} \nonumber \\
& &\times \biggl\{\biggl[\sum _{i=0}^{N}|g_i|^2\biggr]
+2n\biggr[\sum _{i=0}^{N}|g_i|^2
{\rm \delta }\biggl({\bf k}\in {\cal D}(k_0,\sigma _i)\biggr)\biggr]\biggr\}.
\end{eqnarray} 
Our aim is to calculate $G(k) \equiv \sum _{i=0}^N|g_i|^2{\rm \delta
}({\bf k}\in {\cal D}_i)$ [for convenience ${\cal D}(k_0,\sigma _i)$
is denoted by ${\cal D}_i$]. By symmetry, $G(k_0-k')=G(k_0+k')$, so
that we will consider the case $k \equiv k_0 + k'$, $k'\ge 0$. In this
sum $k'$ is fixed. As a consequence, there exists an integer $i_0$
such that if $i < i_0$, ${\rm \delta }({\bf k}\in {\cal D}_i)=0$ and
if $i\ge i_0$, then ${\rm \delta }({\bf k}\in {\cal D}_i)=1$, or,
equivalenty, $\sigma_{i_0} < k'\le \sigma_{i_0+1}$.  This means that
the sum $\sum _{i=0}^N|g_i|^2{\rm \delta }({\bf k}\in {\cal D}_i)$ is
in fact equal to $\sum _{i=i_0}^N|g_i|^2$. We choose the $\sigma_i$'s
and the coefficients $g_i$ according to:
\begin{eqnarray}
\label{defgi}
\sigma_i \equiv i\frac{X_{\rm max}}{N}\quad,\quad
|g_i|^2 \equiv - \frac{X_{\rm max}}{N} \times 
\left.\frac{{\rm d}F}{{\rm d}x} \right|_{x=\sigma_i},
\end{eqnarray}
where $F$ is any decreasing function such that $F(X_{\rm
max})=0$. Then we have
\begin{equation}
\label{defgi2}
\sum_{i=0}^N|g_i|^2{\rm \delta }({\bf k}\in {\cal D}_i)
 = - \frac{X_{\rm max}}{N} 
   \sum _{i=i_0}^N F'\left(i\frac{X_{\rm max}}{N}\right).
\end{equation}
The last step is to send $N$ to infinity. This means that we consider
a continuous series of intervals ${\cal D}_i$. We obtain:
\begin{eqnarray}
\label{Riemansum}
\lim _{N\rightarrow +\infty }
\sum _{i=0}^N|g_i|^2{\rm \delta }({\bf k}\in {\cal D}_i)
 & = & - \int_{k'}^{X_{\rm max}} F'(x){\rm d}x \nonumber \\
 & = & F(k'),
\end{eqnarray}
by definition of the Riemann integral. In the same manner, $\sum
_{i=0}^N|g_i|^2=F(0)$. In what follows, we take
\begin{equation}
\label{defF}
G(k) = F(k')\equiv e^{-\frac{(k-k_0)^2}{\Sigma ^2}},
\end{equation}
where $\Sigma $ is a free parameter. This function does not exactly satisfy
the assumptions made previously, but it is easy to show that
the final result is free of these limitations and is in fact valid for
any function $F$. It is clear that other functions are possible, but
only the approximate shape of the distribution is important and the
Gaussian is the prototype of the function we have in mind. The
spectrum is obtained after having introduced the previous result in
Eq.~(\ref{Kpsi2}) and having taken the limit $L\rightarrow
+\infty$. We obtain:
\begin{equation}
\label{Ppsi2}
k^3P_{\Phi }(k;|\Psi _2 \rangle )=\frac{l_{\rm Pl}^2}{2\pi ^2}
k^3|f_k|^2\biggl(1 +2ne^{-\frac{(k-k_0)^2}{\Sigma ^2}}\biggr).
\end{equation}
In this equation, it is clear that $n$ is an integer. We now show that
this condition can be relaxed if the system is placed in the state
$|\Psi _3 \rangle $.
\par
To calculate the spectrum for this state, it is sufficient to notice
that $\langle \Psi _2(k_0,n)|\Psi _2(k_0,m)\rangle =\delta
_{mn}$. Using this formula, straightforward calculations lead to:
\begin{equation}
\label{Ppsi3}
k^3P_{\Phi }(k;|\Psi _3 \rangle )=\frac{l_{\rm Pl}^2}{2\pi ^2}
k^3|f_k|^2\biggl(1 +2n_{\rm eff}e^{-\frac{(k-k_0)^2}{\Sigma ^2}}\biggr),
\end{equation}
where the effective number of quanta, $n_{\rm eff}$, is given by:
\begin{equation}
\label{defneff}
n_{\rm eff}=\frac{\sum _{n=0}^{\infty }n|h(n)|^2}
                 {\sum _{n=0}^{\infty }|h(n)|^2}.
\label{neff}
\end{equation}
An attractive choice for the function $h(n)$ is obviously $h(n)\equiv
e^{-\beta n}$ [this $\beta$ has of course nothing to do with the
$\beta$ defined in Eq.~(\ref{defpl})]. In this case, $n_{\rm eff}$ is
given by:
\begin{equation}
\label{thermalneff}
n_{\rm eff}=\frac{e^{-2\beta }}{1-e^{-2\beta }}.
\end{equation}
The spectra of Eqs.~(\ref{Ppsi2}) and (\ref{Ppsi3}) are the main
results of this section. Clearly, they possess a peak around the scale
$k_0$. The position of the peak is controlled by the value of $k_0$,
its width by $\Sigma$ and its height by $n$ or $n_{\rm eff}$ (in fact
by $\beta $ in the last case).
\par
We will need the primordial spectrum only for large wavelengths. In the
case of power law inflation, everything can be calculated exactly. In
this limit, we have
\begin{equation}
\label{An0}
k^3P_{\Phi}(k;|0\rangle )=A_{\rm S}k^{n_{\rm S}-1}~,
\end{equation}
with
\begin{equation}
\label{An1}
A_{\rm S}=\frac{l_{\rm Pl}^2}{l_0^2}\frac{\gamma (1+\beta
)^2}{2^{2\beta +4}
\cos ^2(\beta \pi )\Gamma ^2(\beta +5/2)}, \quad n_{\rm S}=2\beta +5.
\end{equation}
The above expression is strictly speaking not applicable in the case of a
de Sitter universe, since then there are no scalar metric perturbations and
the function $\gamma(\eta)$ turns out to be zero. The generation of density 
perturbations is only possible after the transition from the exponential 
inflationary era to the radiation-dominated Universe. If during inflation the 
Universe was very close to the de Sitter space-time, then the spectrum of 
density perturbations today is the so-called Harrison-Zel'dovich spectrum 
($n_{\rm S}=1$). All expressions derived in this section are still valid for 
$\beta {\raise 0.4ex\hbox{$<$}\kern -0.8em\lower 0.62
ex\hbox{$\sim$}} -2$. 
The initial power spectrum in the case where the Bardeen
operator is placed in the state $|\Psi _2\rangle $ can be written as:
\begin{equation}
\label{Ppsi3lwl}
k^3P_{\Phi}(k;|\Psi _2\rangle )=
A_{\rm S}k^{n_{\rm S}-1}
\left(1+2n e^{-\frac{(k-k_0)^2}{\Sigma ^2}}\right).
\end{equation}
If the state is $|\Psi _3\rangle$, we just have to replace the integer
$n$ with the real number $n_{\rm eff}$. Let us note that if, instead
of considering intervals of the form $[k_0-\sigma ,k_0+\sigma ]$, one
considers intervals such as  $]0,k_0+\sigma ]$ or $[k_0-\sigma ,\infty[$,
which still privilege a scale, it is possible to build step-like
spectra, the step being located at the scale $k_0$.
\par
Recently in the literature,  BSI spectra have
also been studied, see Ref.~\cite{bsi}. In these articles, the
privileged scale arises as a privileged energy in the inflaton
potential (more precisely, a discontinuity, or a rapid variation, in the
inflaton potential if first derivatives are present). We would like to
stress that, in our case, the different physical origin of the
privileged scale would in principle allow us to distinguish the different
models. Indeed, in Ref.~\cite{bsi}, the fluctuations are Gaussian. In
the case studied here, the three-point correlation function still
vanishes
\begin{equation}
\label{3points}
\left\langle{{\rm \delta} T\over T}({\bf e}_1)
            {{\rm \delta} T\over T}({\bf e}_2)
            {{\rm \delta} T\over T}({\bf e}_3)\right\rangle = 0 \;~,
\end{equation}
but the four-point correlation function no longer satisfies the
relation
\begin{equation}
\label{4points}
\left\langle\biggl({{\rm \delta} T\over T}({\bf e})\biggr)^4\right\rangle 
= 3 \Biggl[\left\langle\biggl({{\rm \delta} T\over T}({\bf
e})\biggr)^2\right\rangle\Biggr] ^2,
\end{equation}
which is typical of Gaussian statistics. The reason for this is
clear. The ground-state wave function of an harmonic oscillator is a
Gaussian and, as a consequence, the CMBR correlation functions for the
vacuum exhibit Gaussian properties. On the other hand, the wave
function of a state with a non-vanishing number of quanta is no longer
a Gaussian and, correspondingly, the correlation functions deviate from
Gaussianity. Therefore, a measure of the four-point correlation
function (as well as any higher-order even-point correlation
function) would permit to distinguish between the class of models
presented here and the models of Ref.~\cite{bsi}. If it turns out that
the type of non-Gaussianity apparently detected recently
\cite{non-gaus1,non-gaus2,non-gaus3} is really present in the CMBR
map, then these two classes of BSI models (as well as standard
inflation) are ruled out, because they both predict a vanishing
three-point correlation function. But if it turns out that some 
non-Gaussianity is present in the CMBR at the level of the four-point
correlation function then  the models presented 
here could account for this.

\section{Comparison with observations}

The aim of this section is to confront the power spectra given by
Eq.~(\ref{Ppsi3lwl}) with observations. We will not use any accurate
statistical methods to find the best values of the free parameters
$k_0$, $\Sigma $ and $n$/$n_{\rm eff}$, because we just want to obtain
crude constraints.  For this purpose we will use observations of the
CMBR anisotropies and of the matter power spectrum.
\par
We choose to work with the following cosmological parameters: the
Hubble parameter is $h=0.5$, the baryonic matter-density parameter is
$\Omega _b=0.05$, the density parameter $\Omega_0 \equiv
\Omega _c+\Omega _b+\Omega_\Lambda$ is equal to 1 ($\Omega_c$
and $\Omega_\Lambda$ are respectively the CDM and the $\Lambda$-density 
parameters), there is no significant reionization and the spectral 
index is $n_{\rm S}=1$, when 
there is no contribution from the gravitational waves.
 Let us emphasize again that in this case, since it corresponds to a de 
Sitter phase,  the  Eq.~(\ref{An1}) giving the normalization 
of the spectrum is strictly speaking not applicable. 
However, all expressions derived in the previous section can be applied for
$n_{\rm S}  {\raise 0.4ex\hbox{$<$}\kern -0.8em\lower 0.62
ex\hbox{$\sim$}} 1$. We will later discuss the case of small deviations 
from a scale-invariant (Harrison-Zel'dovich) spectrum, including the
contribution of gravitational waves in the CMBR anisotropies. 
We will consider 
two different values for
the cosmological constant density parameter $\Omega _{\Lambda }\equiv
\Lambda/(3H_0^2)$ and the sum of baryon-matter density parameter and
CDM density parameter $\Omega _m\equiv \Omega _b+\Omega
_c$, that is $\Omega_\Lambda=0$, $\Omega_c = 0.95$ (hereafter denoted
SCDM), and $\Omega_\Lambda=0.6$, $\Omega_c = 0.35$ (hereafter denoted
$\Lambda$CDM).
 We point out that we have not assumed any biasing in the galaxy distribution
with respect to the underlying mass fluctuations (the bias parameter is equal to 1).

\par
The spectrum must be normalized, i.e. the value of $A_{\rm S}$ must be
determined. For this purpose, we use the value of $Q_{\rm
rms-PS}=T_0(5C_2/4\pi )^{1/2} \sim 18 ~{\rm \mu K}$ ($T_0=2.7 ~{\rm K}$)
measured by the COBE satellite. We use the value $Q_{\rm rms-PS}\sim
18 ~{\rm \mu K}$ because we have assumed that the spectrum is 
scale-invariant. In the large-wavelength approximation, we have ${\rm \delta
}T/T\sim (1/3)\Phi $. In addition the transfer function  for the Bardeen 
potential can be taken equal to 1 (with an appropriate 
normalization). As a consequence the
multipole can be written as:
\begin{eqnarray}
\label{cl}
C_\ell & = & \frac{4\pi }{9}\int _0^{+\infty }\frac{{\rm d}k}{k}
\biggl[j_\ell[k(\eta_0-\eta_{\rm LSS})]^2 \nonumber \\
 & & \quad \times 
A_{\rm S}(n_{\rm S})k^{n_{\rm S}-1}
\left(1+2ne^{-\frac{(k-k_0)^2}{\Sigma ^2}}\right)
\biggr],
\end{eqnarray}
where $j_\ell$ is a spherical Bessel function of order $\ell$, and
$\eta_0$ and $\eta_{\rm LSS}$ denote respectively the conformal times
now and at the last scattering surface. Let us remark that the $A_{\rm
S}$ in the last expression is not exactly the $A_{\rm S}$ in
Eqs.~(\ref{An0}) and (\ref{An1}). Since the difference is not important
for our purpose, we have kept the same notation. The previous
expression can be evaluated explicitly. For the quadrupole, the result
reads:
\begin{eqnarray}
\label{quadrupole}
 & & C_2 =\frac{4 \pi}{9} 
     A_{\rm S}(n_{\rm S})(\eta_0-\eta_{\rm LSS})^{1-n_{\rm S}}
\nonumber \\ & & \times \biggl(\frac{\pi}{2^{4-n_{\rm S}}}
\frac{\Gamma [3-n_{\rm S}]\Gamma [2+(n_{\rm S}-1)/2]}
     {\Gamma ^2[(4-n_{\rm S})/2]\Gamma [4-(n_{\rm S}-1)/2]}
+2n I \biggr) ,
\end{eqnarray}
where 
\begin{equation}
\label{I}
I\equiv \int _0^{+\infty }
\frac{{\rm d}u}{u^{2-n_s}}
[j_2(u)]^2
e^{-\frac{(u-u_0)^2}{U^2}},
\end{equation}
with $u_0 \equiv k_0(\eta_0-\eta_{\rm LSS})$ and $U \equiv
\Sigma(\eta_0-\eta_{\rm LSS})$. In what follows, we take 
$\eta _0-\eta _{LSS}=1$. The integral will be evaluated numerically
for different values of the free parameters. We just have to
specialize the last equation to a scale-invariant spectrum to obtain
the following value for $A_{\rm S}$:
\begin{equation}
\label{A_S}
A_{\rm S}=\frac{108}{5T_0^2}Q^2_{\rm rms-PS}\frac{1}{1+24 n I}= 
\frac{9.4\cdot 10^{-10}}{1+24 n I}.
\end{equation}
In terms of the band power ${\rm \delta }T_\ell$ defined by ${\rm
\delta }T_\ell\equiv T_0[\ell(\ell+1)C_\ell/2\pi]^{1/2}$, we find 
${\rm \delta }T_2=\sqrt{12/5}Q_{\rm rms-PS}=27.9 ~{\rm \mu K}$.
\par
\begin{figure}
\begin{center}
\leavevmode
\begingroup%
  \makeatletter%
  \newcommand{\GNUPLOTspecial}{%
    \@sanitize\catcode`\%=14\relax\special}%
  \setlength{\unitlength}{0.06bp}%
\begin{picture}(3600,2000)(0,0)%
\special{psfile=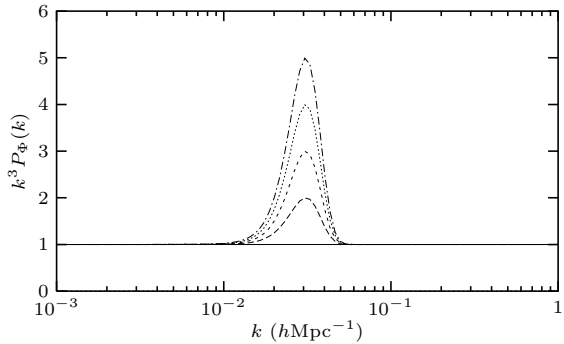 llx=0 lly=0 urx=720 ury=504 rwi=4320}
\put(1875,50){\makebox(0,0){\scriptsize  $k$ ($h$Mpc$^{-1}$)}}%
\put(100,1180){%
\special{ps: gsave currentpoint currentpoint translate
270 rotate neg exch neg exch translate}%
\makebox(0,0)[b]{\shortstack{\scriptsize $k^3 P_\Phi (k)$}}%
\special{ps: currentpoint grestore moveto}%
}%
\put(3450,200){\makebox(0,0){\scriptsize 1}}%
\put(2400,200){\makebox(0,0){\scriptsize $10^{-1}$}}%
\put(1350,200){\makebox(0,0){\scriptsize $10^{-2}$}}%
\put(300,200){\makebox(0,0){\scriptsize $10^{-3}$}}%
\put(250,2060){\makebox(0,0)[r]{\scriptsize 6}}%
\put(250,1767){\makebox(0,0)[r]{\scriptsize 5}}%
\put(250,1473){\makebox(0,0)[r]{\scriptsize 4}}%
\put(250,1180){\makebox(0,0)[r]{\scriptsize 3}}%
\put(250,887){\makebox(0,0)[r]{\scriptsize 2}}%
\put(250,593){\makebox(0,0)[r]{\scriptsize 1}}%
\put(250,300){\makebox(0,0)[r]{\scriptsize 0}}%
\end{picture}%
\endgroup
 
\end{center}
\caption{Initial power spectrum for $n_{\rm eff}$ ranging from 0 to
2 with steps of 0.5. Vertical units are arbitrary.}
\label{ps_1}
\end{figure}

We must choose the three parameters $k_0$, $n$ and $\Sigma
$. Recently, it has been emphasized by many authors \cite{einasto}
 that the power
spectrum seems to contain large amplitude features at the scale
$l_{\rm C}\approx 100 ~h^{-1}~\Mpc$, which corresponds to a wave number
equal to $0.062 ~h~\Mpc^{-1}$. No other value for a privileged scale has
been detected so far, and therefore any other choice would either lie
in an unobservable range, or be in conflict with the data available at
present. Consequently, we choose:
\begin{equation}
\label{position}
k_0=0.062 ~h~ \Mpc^{-1}=0.031 ~\Mpc^{-1},
\end{equation}
with our value of the Hubble constant. Let us turn to the choice of
the variance $\Sigma$. We have seen that the  simplest non-vacuum
initial states can lead to a power spectrum with either a bump or a
step. In this article, we will restrict ourselves to the study of the
bump case. Step-like spectra have already been studied in
Ref.~\cite{bsi} and our conclusions would be similar. Therefore we
will consider (rather arbitrarily, but the conclusion does not depend on
the exact value of $\Sigma$, as long as it is not too large):
\begin{equation}
\label{variance}
\Sigma =0.3k_0=0.0186 ~h~ \Mpc^{-1} .
\end{equation}
>From now on, we will always take these two values for $k_0$ and
$\Sigma$ in any of the plots shown.  In Fig.~\ref{ps_1} we display the
initial power spectrum for a few values of $n_{\rm eff}$. The
difference between $n_{\rm eff}=0$ and $n_{\rm eff}\neq 0$ is obvious.
In the case considered here, the integral $I$ is equal to: $I(\Sigma
=0.3k_0)\approx 1.3 \cdot 10^{-6}$. It is completely negligible and
will be taken equal to zero. This arises from the fact that the
quadrupole is mainly fed by very large wavelengths (of the order of today's
Hubble radius), whereas the bump occurs at much smaller wavelengths
(of the order of the Hubble radius at the time of decoupling). Thus, the
calculation of the quadrupole, and therefore the normalization, is not
modified by the presence of the bump.
\par
Let us discuss the matter power spectrum. The power spectrum can
either be obtained by the Boltzmann code developed by one of us
(A.~R.) or by means of analytical fits. In this case, 
the  baryons power spectrum is given by:
\begin{equation}
\label{psmatter}
\frac{{\rm \delta }\rho _{\rm b}}{\rho _{\rm b}}\equiv |\delta (k)|^2=AT^2(k)
\frac{g^2(\Omega _0)}{g^2(\Omega_m)}k\biggl[1+2ne^{-\frac{(k-k_0)^2}
{\Sigma ^2}}\biggr],
\end{equation}
where the different terms in this equation are explained below; $T(k)$
is the transfer function, which can be approximated by the following
numerical fit~\cite{bbks}:
\begin{eqnarray}
\label{transfert}
T(k) & = & \frac{\ln (1+2.34 q)}{2.34 q}[1+3.89q \nonumber \\
& & +(16.1q)^2+(5.46q)^3+(6.71q)^4]^{-1/4},
\end{eqnarray}
with $q\equiv k/[(h\Gamma )\Mpc^{-1}]$ where $\Gamma $ is the
so-called shape parameter, which can be written as~\cite{sugiyama}:
\begin{equation}
\label{shapepara}
\Gamma \equiv \Omega _mhe^{-\Omega _b-\frac{\Omega _b}{\Omega _m}}.
\end{equation}
The function $g(\Omega)$ takes into account the modification induced
in the power spectrum by the presence of a cosmological constant. Its
expression can be written as~\cite{primack}:
\begin{equation}
\label{functiong}
g(\Omega )\equiv \frac{5\Omega }{2}\biggl[\Omega ^{4/7}-\Omega _{\Lambda }+
\biggl(1+\frac{\Omega }{2}\biggr)\biggl(1+\frac{\Omega _{\Lambda }}{70}\biggr)
\biggr]^{-1}.
\end{equation}
Finally the coefficient $A$ is the normalization. We normalize
the spectrum to COBE data. This leads to the following value for $A$:
\begin{eqnarray}
\label{A}
A &=& (2l_H)^4\frac{6\pi ^2}{5}
\frac{Q_{\rm rms-PS}^2}{T_0^2}\frac{1}{1+24 n I} \\
&=& \frac{6.82\cdot 10^5}{1+24 nI}h^{-4}\Mpc^4,
\end{eqnarray}
where the Hubble radius, $l_H$, is equal to $3000h^{-1}\Mpc$. 
\par
We plot the multipole moments and the power spectrum for different
values of $n$ and/or $n_{\rm eff}$. The $C_\ell$'s are obtained from
the Boltzmann code previously used for the power spectrum.  In all
figures for the $C_\ell$'s, we represent the COBE data~\cite{FIRS} by
diamonds, the Saskatoon data~\cite{Saskatoon} by squares, and the
CAT~\cite{CAT} data by crosses. (For clarity we have not displayed all
CMBR data on the figures.) In all figures for the power spectra, we
represent the APM data~\cite{APM} by diamonds, the velocities field
measurements~\cite{vel} by squares, and the data given by Einasto {\it
et al.}~\cite{einasto} by crosses. 

\subsection{Scalar modes only}

\begin{figure}
\begin{center}
\leavevmode
\begingroup%
  \makeatletter%
  \newcommand{\GNUPLOTspecial}{%
    \@sanitize\catcode`\%=14\relax\special}%
  \setlength{\unitlength}{0.065bp}%
\begin{picture}(3600,2000)(0,0)%
\special{psfile=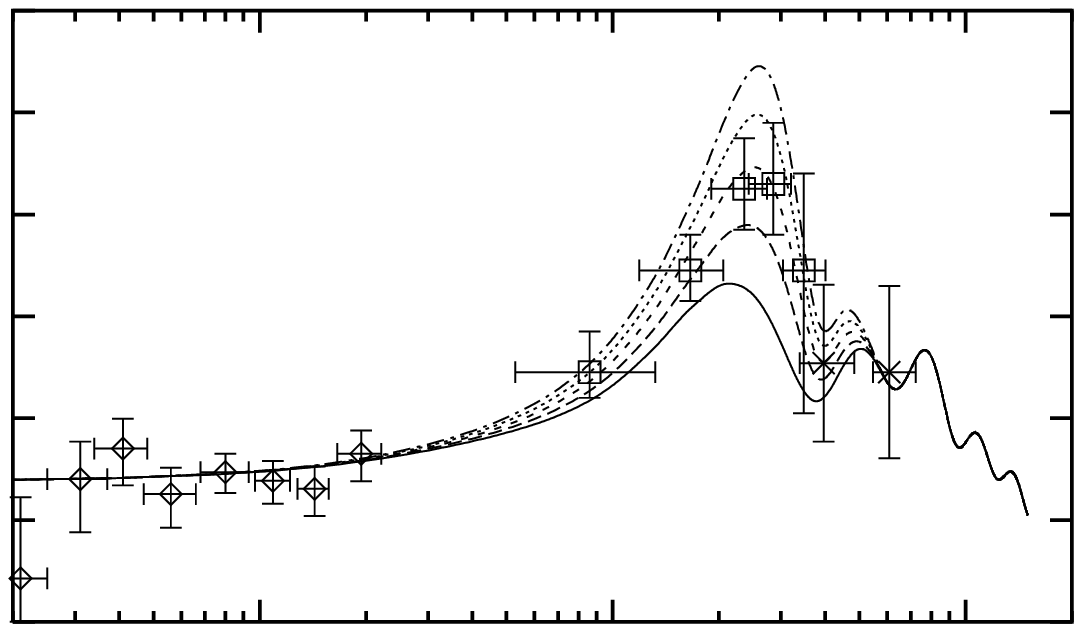 llx=0 lly=0 urx=720 ury=504 rwi=4680}
\put(1925,50){\makebox(0,0){\scriptsize  $\ell$}}%
\put(100,1180){%
\special{ps: gsave currentpoint currentpoint translate
270 rotate neg exch neg exch translate}%
\makebox(0,0)[b]{\shortstack{\scriptsize $T_0 [\ell(\ell+1)C_\ell /(2\pi)]^{1/2}$ ($\mu$K)}}%
\special{ps: currentpoint grestore moveto}%
}%
\put(3144,200){\makebox(0,0){\scriptsize 1000}}%
\put(2127,200){\makebox(0,0){\scriptsize 100}}%
\put(1111,200){\makebox(0,0){\scriptsize 10}}%
\put(350,2060){\makebox(0,0)[r]{\scriptsize 120}}%
\put(350,1767){\makebox(0,0)[r]{\scriptsize 100}}%
\put(350,1473){\makebox(0,0)[r]{\scriptsize 80}}%
\put(350,1180){\makebox(0,0)[r]{\scriptsize 60}}%
\put(350,887){\makebox(0,0)[r]{\scriptsize 40}}%
\put(350,593){\makebox(0,0)[r]{\scriptsize 20}}%
\put(350,300){\makebox(0,0)[r]{\scriptsize 0}}%
\end{picture}%
\endgroup
 
\end{center}
\caption{Multipole moments for the SCDM model with $n_{\rm eff}$ (and
$n$ if it is integer) ranging from 0 to 2 with step of 0.5 (from the
bottom to the top).  Diamonds represent COBE data, squares the
Saskatoon data, and crosses the CAT data.}
\label{cl_1}
\end{figure}

We first display the CMBR anisotropies in the SCDM model
(Fig.~\ref{cl_1}).
In the case were $\Lambda=0$, Saskatoon data are compatible with the
case $n_{\rm eff}=1$ (third curve). 

\begin{figure}
\begin{center}
\leavevmode
\begingroup%
  \makeatletter%
  \newcommand{\GNUPLOTspecial}{%
    \@sanitize\catcode`\%=14\relax\special}%
  \setlength{\unitlength}{0.065bp}%
\begin{picture}(3600,2000)(0,0)%
\special{psfile=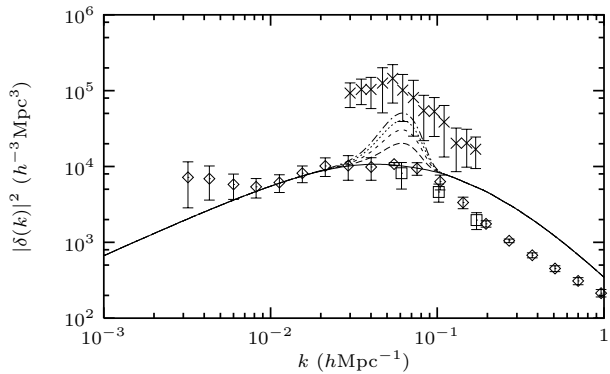 llx=0 lly=0 urx=720 ury=504 rwi=4680}
\put(2000,50){\makebox(0,0){\scriptsize  $k$ ($h$Mpc$^{-1}$)}}%
\put(100,1180){%
\special{ps: gsave currentpoint currentpoint translate
270 rotate neg exch neg exch translate}%
\makebox(0,0)[b]{\shortstack{\scriptsize $|\delta(k)|^2$ ($h^{-3}$Mpc$^3$)}}%
\special{ps: currentpoint grestore moveto}%
}%
\put(3450,200){\makebox(0,0){\scriptsize 1}}%
\put(2483,200){\makebox(0,0){\scriptsize $10^{-1}$}}%
\put(1517,200){\makebox(0,0){\scriptsize $10^{-2}$}}%
\put(550,200){\makebox(0,0){\scriptsize $10^{-3}$}}%
\put(500,2060){\makebox(0,0)[r]{\scriptsize $10^6$}}%
\put(500,1620){\makebox(0,0)[r]{\scriptsize $10^5$}}%
\put(500,1180){\makebox(0,0)[r]{\scriptsize $10^4$}}%
\put(500,740){\makebox(0,0)[r]{\scriptsize $10^3$}}%
\put(500,300){\makebox(0,0)[r]{\scriptsize $10^2$}}%
\end{picture}%
\endgroup
 
\end{center}
\caption{Power spectrum for the SCDM model, with $n_{\rm eff}$ ranging
from 0 to 2 with step of 0.5 (from the bottom to the top). Diamonds
represent the APM data, squares the velocities field
measurements, and crosses  the data  by Einasto {\it et
al.} }
\label{ps_2}
\end{figure}

We note that the position of the first Doppler peak is no longer
around $\ell \approx 220$. Usually, its position is determined by the
angular size of the Hubble radius at recombination. In our case, we
must superimpose the bump present in the initial spectrum, the
position of which is not at $\ell \approx 220$ but rather at the
angular scale sustained by the built-in scale. As a consequence, the
resulting peak is shifted towards higher values of $\ell$ for the
values of the parameters considered here ($\ell \approx 260$). In
addition, it could be difficult to distinguish the effect due to the
primordial bump from the one coming from a variation of the
cosmological parameters, thus increasing the degeneracy among the free
parameters of the model. Let us note, however, that the bump in the
initial power spectrum, should be easier to detect in the matter power
spectrum since it is a more slowly varying function, as shown in
Fig.~\ref{ps_2}.
A higher value of $n_{\rm eff}$ (2 rather than 1) seems to be needed
to explain the data of Einasto {\it et al.}, but different cosmological 
parameters might
lead to a better agreement between CMBR and matter power spectrum
data.
\par
\begin{figure}
\begin{center}
\leavevmode
\begingroup%
  \makeatletter%
  \newcommand{\GNUPLOTspecial}{%
    \@sanitize\catcode`\%=14\relax\special}%
  \setlength{\unitlength}{0.065bp}%
\begin{picture}(3600,2000)(0,0)%
\special{psfile=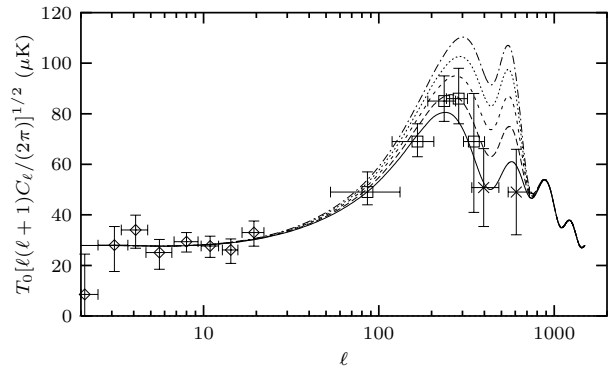 llx=0 lly=0 urx=720 ury=504 rwi=4680}
\put(1925,50){\makebox(0,0){\scriptsize  $\ell$}}%
\put(100,1180){%
\special{ps: gsave currentpoint currentpoint translate
270 rotate neg exch neg exch translate}%
\makebox(0,0)[b]{\shortstack{\scriptsize $T_0 [\ell(\ell+1)C_\ell /(2\pi)]^{1/2}$ ($\mu$K)}}%
\special{ps: currentpoint grestore moveto}%
}%
\put(3144,200){\makebox(0,0){\scriptsize 1000}}%
\put(2127,200){\makebox(0,0){\scriptsize 100}}%
\put(1111,200){\makebox(0,0){\scriptsize 10}}%
\put(350,2060){\makebox(0,0)[r]{\scriptsize 120}}%
\put(350,1767){\makebox(0,0)[r]{\scriptsize 100}}%
\put(350,1473){\makebox(0,0)[r]{\scriptsize 80}}%
\put(350,1180){\makebox(0,0)[r]{\scriptsize 60}}%
\put(350,887){\makebox(0,0)[r]{\scriptsize 40}}%
\put(350,593){\makebox(0,0)[r]{\scriptsize 20}}%
\put(350,300){\makebox(0,0)[r]{\scriptsize 0}}%
\end{picture}%
\endgroup
 
\end{center}
\caption{Same as Fig.~\ref{cl_1}, but for the $\Lambda$CDM model, with
$\Omega_\Lambda = 0.6$.}
\label{cl_2}
\end{figure}
\begin{figure}
\begin{center}
\leavevmode
\begingroup%
  \makeatletter%
  \newcommand{\GNUPLOTspecial}{%
    \@sanitize\catcode`\%=14\relax\special}%
  \setlength{\unitlength}{0.065bp}%
\begin{picture}(3600,2000)(0,0)%
\special{psfile=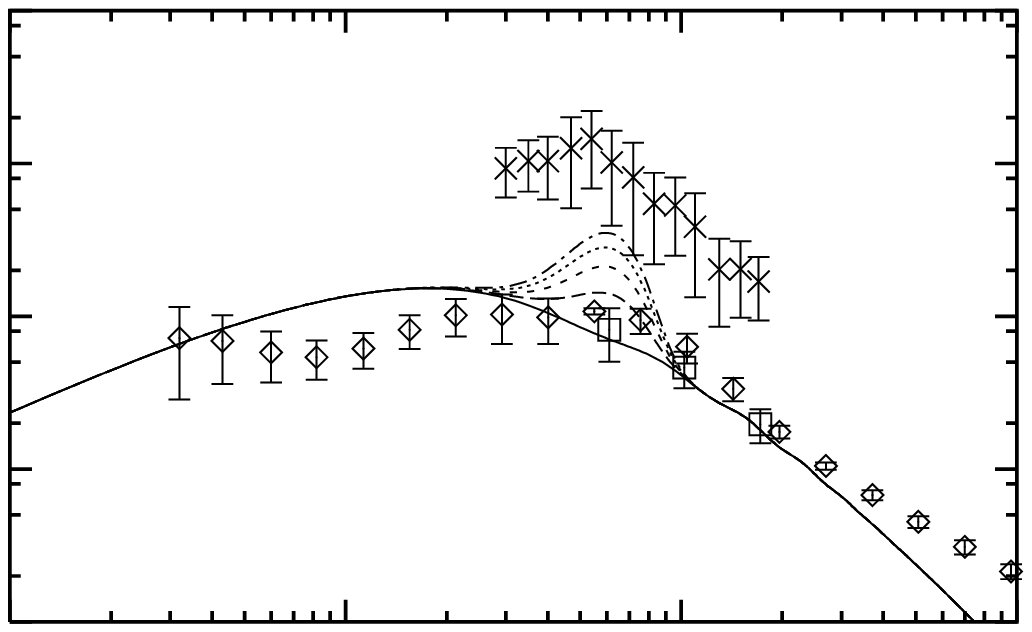 llx=0 lly=0 urx=720 ury=504 rwi=4680}
\put(2000,50){\makebox(0,0){\scriptsize  $k$ ($h$Mpc$^{-1}$)}}%
\put(100,1180){%
\special{ps: gsave currentpoint currentpoint translate
270 rotate neg exch neg exch translate}%
\makebox(0,0)[b]{\shortstack{\scriptsize $|\delta(k)|^2$ ($h^{-3}$Mpc$^3$)}}%
\special{ps: currentpoint grestore moveto}%
}%
\put(3450,200){\makebox(0,0){\scriptsize 1}}%
\put(2483,200){\makebox(0,0){\scriptsize $10^{-1}$}}%
\put(1517,200){\makebox(0,0){\scriptsize $10^{-2}$}}%
\put(550,200){\makebox(0,0){\scriptsize $10^{-3}$}}%
\put(500,2060){\makebox(0,0)[r]{\scriptsize $10^6$}}%
\put(500,1620){\makebox(0,0)[r]{\scriptsize $10^5$}}%
\put(500,1180){\makebox(0,0)[r]{\scriptsize $10^4$}}%
\put(500,740){\makebox(0,0)[r]{\scriptsize $10^3$}}%
\put(500,300){\makebox(0,0)[r]{\scriptsize $10^2$}}%
\end{picture}%
\endgroup
 
\end{center}
\caption{Same as Fig.~\ref{ps_2}, but for the $\Lambda$CDM model.}
\label{ps_3}
\end{figure}

We now display the CMBR (Fig.~\ref{cl_2}) and matter power spectrum
(Fig.~\ref{ps_3}) of the $\Lambda$CDM model.
When the cosmological constant $\Omega_\Lambda = 0.6$, the early Integrated 
Sachs--Wolfe effect already boosts the $\ell\simeq 200-300$ scale 
sufficiently \cite{Hu_1}: at $n_{\rm eff}=1$ this effect already puts too
much power on these scales. A 
different value for $k_0$ and $\Sigma$ might also be needed 
to remain compatible with the CAT data. For the matter power 
spectrum, the same conclusion as for the SCDM model holds, that is a
higher value of $n_{\rm eff}$ is preferred (around 2 or 3).
\par
As a conclusion of this rapid analysis, we stress that our model is
much more constrained if one imposes $n_{\rm eff}$ to be an integer
instead of a real number. Moreover, our model tends to favour a moderate
value of $n_{\rm eff}$ as well as a low value of the cosmological
constant if the data of Einasto {\it et al.} are confirmed, or a 
low value of $n_{\rm eff}$ and a high value of the cosmological constant 
(i.e.~the currently popular cosmological model, with vacuum initial state) in
the other case. It is easy to notice from Eq.~(\ref{neff}), 
that since $n_{\rm eff}$ is quite small, $h(n)$ is peaked around  small 
values of $n$, and therefore the allowed window for the effective number of 
quanta is constrained to be around small values. In conclusion,
the initial state found is not too far from the vacuum.

\subsection{Scalar and tensor modes}

\begin{figure}
\begin{center}
\leavevmode
\begingroup%
  \makeatletter%
  \newcommand{\GNUPLOTspecial}{%
    \@sanitize\catcode`\%=14\relax\special}%
  \setlength{\unitlength}{0.065bp}%
\begin{picture}(3600,2000)(0,0)%
\special{psfile=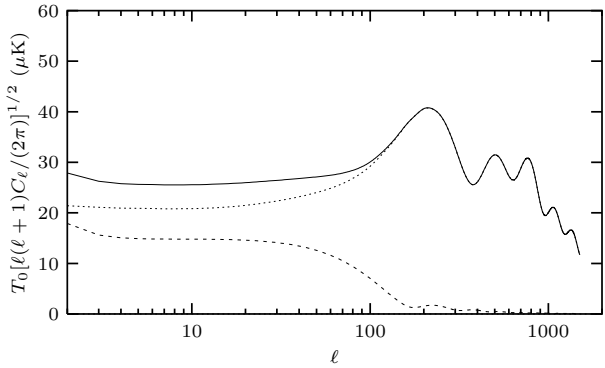 llx=0 lly=0 urx=720 ury=504 rwi=4680}
\put(1900,50){\makebox(0,0){\scriptsize  $\ell$}}%
\put(100,1180){%
\special{ps: gsave currentpoint currentpoint translate
270 rotate neg exch neg exch translate}%
\makebox(0,0)[b]{\shortstack{\scriptsize $T_0 [\ell(\ell+1)C_\ell /(2\pi)]^{1/2}$ ($\mu$K)}}%
\special{ps: currentpoint grestore moveto}%
}%
\put(3139,200){\makebox(0,0){\scriptsize 1000}}%
\put(2106,200){\makebox(0,0){\scriptsize 100}}%
\put(1072,200){\makebox(0,0){\scriptsize 10}}%
\put(300,2060){\makebox(0,0)[r]{\scriptsize 60}}%
\put(300,1767){\makebox(0,0)[r]{\scriptsize 50}}%
\put(300,1473){\makebox(0,0)[r]{\scriptsize 40}}%
\put(300,1180){\makebox(0,0)[r]{\scriptsize 30}}%
\put(300,887){\makebox(0,0)[r]{\scriptsize 20}}%
\put(300,593){\makebox(0,0)[r]{\scriptsize 10}}%
\put(300,300){\makebox(0,0)[r]{\scriptsize 0}}%
\end{picture}%
\endgroup
 
\end{center}
\caption{CMBR anisotropies decomposition, showing scalar (dotted
line) and tensor (dashed line) contributions. The total contribution
is given by the solid line.}
\label{cl_3}
\end{figure}

One should also consider the contribution of the gravitational waves
in the CMBR anisotropies. The data currently available are in fact the
sum of the scalar plus the tensor contributions to the CMBR
anisotropies. We recall that since there are two modes of polarization
for the CMBR photons and that one of them is only generated by
gravitational waves, it is in principle possible to distinguish
between the scalar and tensor contributions to the CMBR anisotropies,
see~\cite{Hu_2}. In what follows, we consider some standard
inflationary predictions for gravitational waves: we take $n_{\rm S} =
0.9$, $n_{\rm T} = n_{\rm S}-1 \approx -0.1$ (the last equation being
rigorous in the case of power-law inflation only), and the ratio of
scalar to tensor amplitude $C_2^{\rm T} / C_2^{\rm S}
\approx - 7 n_{\rm T}$. In Fig.~\ref{cl_3}, we decompose CMBR
anisotropies, showing the contributions from scalar and
tensor modes separately.
In any model, the gravitational waves contribution can be important only
for multipoles $\ell ~\lsim ~ 100$, while it is negligible at smaller
angular scales (roughly speaking, the gravitational waves contribution
is two orders of magnitude smaller at $\ell \approx 300$ than at the
quadrupole). The effect of gravitational waves is therefore to boost
power on large angular scales (or, equivalently, to lower the height of
the acoustic peaks with respect to the height of the low $\ell$
plateau). The fact that one observes an excess of power on small
angular scales (with Saskatoon data), favours a low contribution from
gravitational waves (which is in agreement with most inflationary
models). In our model, the possibility to have a bump in the initial
power spectrum enables us to boost the height of the acoustic peaks,
and therefore to have some non-negligible contribution from
gravitational waves: normalizing at COBE data, one imposes the value
of $A_{\rm S} + A_{\rm T}$ instead of $A_{\rm S}$. As a result, the
scalar perturbations amplitude $A_{\rm S}$ is smaller. Since the first
acoustic peak depends only on scalar perturbations, we must keep the
same value as before for the product $A_{\rm S}(1+2n\exp[-(k_0-k_{\rm
peak})^2/\Sigma^2])$, which permits a higher value of $n_{\rm eff}$
($k_{\rm peak}$ is the characteric wave number of the first Doppler
peak).
\par
\begin{figure}
\begin{center}
\leavevmode
\begingroup%
  \makeatletter%
  \newcommand{\GNUPLOTspecial}{%
    \@sanitize\catcode`\%=14\relax\special}%
  \setlength{\unitlength}{0.065bp}%
\begin{picture}(3600,2000)(0,0)%
\special{psfile=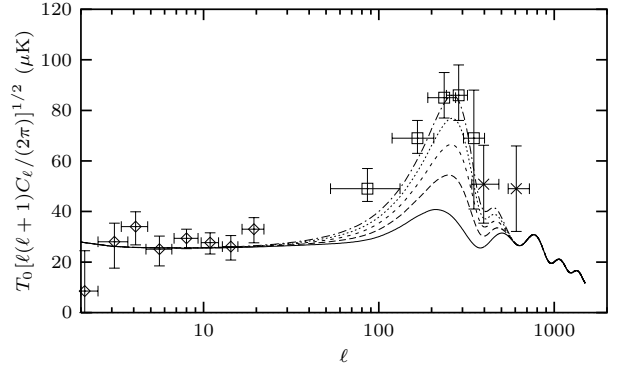 llx=0 lly=0 urx=720 ury=504 rwi=4680}
\put(1925,50){\makebox(0,0){\scriptsize  $\ell$}}%
\put(100,1180){%
\special{ps: gsave currentpoint currentpoint translate
270 rotate neg exch neg exch translate}%
\makebox(0,0)[b]{\shortstack{\scriptsize $T_0 [\ell(\ell+1)C_\ell /(2\pi)]^{1/2}$ ($\mu$K)}}%
\special{ps: currentpoint grestore moveto}%
}%
\put(3144,200){\makebox(0,0){\scriptsize 1000}}%
\put(2127,200){\makebox(0,0){\scriptsize 100}}%
\put(1111,200){\makebox(0,0){\scriptsize 10}}%
\put(350,2060){\makebox(0,0)[r]{\scriptsize 120}}%
\put(350,1767){\makebox(0,0)[r]{\scriptsize 100}}%
\put(350,1473){\makebox(0,0)[r]{\scriptsize 80}}%
\put(350,1180){\makebox(0,0)[r]{\scriptsize 60}}%
\put(350,887){\makebox(0,0)[r]{\scriptsize 40}}%
\put(350,593){\makebox(0,0)[r]{\scriptsize 20}}%
\put(350,300){\makebox(0,0)[r]{\scriptsize 0}}%
\end{picture}%
\endgroup
 
\end{center}
\caption{CMBR anisotropies for the SCDM model, with $n_{\rm eff}$
ranging from 0 to 4 with a step of 1 (from the bottom to the
top). Both scalar and tensor contributions are included. Diamonds
represent COBE data, squares the Saskatoon data, and crosses 
the CAT data.}
\label{cl_4}
\end{figure}
\begin{figure}
\begin{center}
\leavevmode
\begingroup%
  \makeatletter%
  \newcommand{\GNUPLOTspecial}{%
    \@sanitize\catcode`\%=14\relax\special}%
  \setlength{\unitlength}{0.065bp}%
\begin{picture}(3600,2000)(0,0)%
\special{psfile=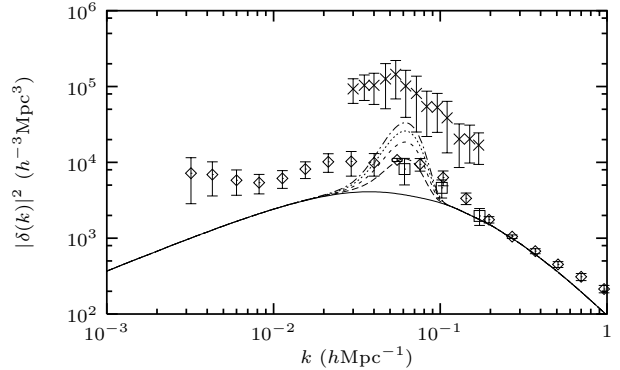 llx=0 lly=0 urx=720 ury=504 rwi=4680}
\put(2000,50){\makebox(0,0){\scriptsize  $k$ ($h$Mpc$^{-1}$)}}%
\put(100,1180){%
\special{ps: gsave currentpoint currentpoint translate
270 rotate neg exch neg exch translate}%
\makebox(0,0)[b]{\shortstack{\scriptsize $|\delta(k)|^2$ ($h^{-3}$Mpc$^3$)}}%
\special{ps: currentpoint grestore moveto}%
}%
\put(3450,200){\makebox(0,0){\scriptsize 1}}%
\put(2483,200){\makebox(0,0){\scriptsize $10^{-1}$}}%
\put(1517,200){\makebox(0,0){\scriptsize $10^{-2}$}}%
\put(550,200){\makebox(0,0){\scriptsize $10^{-3}$}}%
\put(500,2060){\makebox(0,0)[r]{\scriptsize $10^6$}}%
\put(500,1620){\makebox(0,0)[r]{\scriptsize $10^5$}}%
\put(500,1180){\makebox(0,0)[r]{\scriptsize $10^4$}}%
\put(500,740){\makebox(0,0)[r]{\scriptsize $10^3$}}%
\put(500,300){\makebox(0,0)[r]{\scriptsize $10^2$}}%
\end{picture}%
\endgroup
 
\end{center}
\caption{Power spectrum for the SCDM model, with $n_{\rm eff}$ ranging from 
0 to 4 with step of 1. Diamonds represent the  APM data,  squares  the 
velocities field measurements, and crosses  the data given by 
Einasto {{\it et al.}}} 
\label{ps_4}
\end{figure}

In Figs.~\ref{cl_4} and \ref{ps_4} we show the CMBR anisotropies and the
matter power spectrum for the SCDM model, including both scalar and
tensor contributions. 
Comparing these figures with Figs.~\ref{cl_1} and\ref{ps_2}, it can be
concluded that if both scalar and tensor modes are included in the
calculation of the multipole moments $C_\ell$'s, then a higher number
of quanta ($\simeq 4$) is required as expected.
\par
\begin{figure}
\begin{center}
\leavevmode
\begingroup%
  \makeatletter%
  \newcommand{\GNUPLOTspecial}{%
    \@sanitize\catcode`\%=14\relax\special}%
  \setlength{\unitlength}{0.065bp}%
\begin{picture}(3600,2000)(0,0)%
\special{psfile=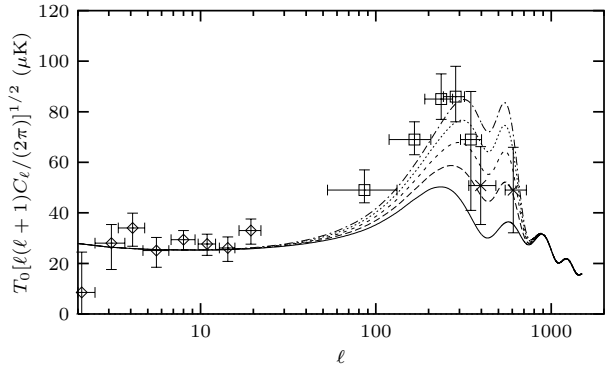 llx=0 lly=0 urx=720 ury=504 rwi=4680}
\put(1925,50){\makebox(0,0){\scriptsize  $\ell$}}%
\put(100,1180){%
\special{ps: gsave currentpoint currentpoint translate
270 rotate neg exch neg exch translate}%
\makebox(0,0)[b]{\shortstack{\scriptsize $T_0 [\ell(\ell+1)C_\ell /(2\pi)]^{1/2}$ ($\mu$K)}}%
\special{ps: currentpoint grestore moveto}%
}%
\put(3144,200){\makebox(0,0){\scriptsize 1000}}%
\put(2127,200){\makebox(0,0){\scriptsize 100}}%
\put(1111,200){\makebox(0,0){\scriptsize 10}}%
\put(350,2060){\makebox(0,0)[r]{\scriptsize 120}}%
\put(350,1767){\makebox(0,0)[r]{\scriptsize 100}}%
\put(350,1473){\makebox(0,0)[r]{\scriptsize 80}}%
\put(350,1180){\makebox(0,0)[r]{\scriptsize 60}}%
\put(350,887){\makebox(0,0)[r]{\scriptsize 40}}%
\put(350,593){\makebox(0,0)[r]{\scriptsize 20}}%
\put(350,300){\makebox(0,0)[r]{\scriptsize 0}}%
\end{picture}%
\endgroup
 
\end{center}
\caption{Same as Fig. \ref{cl_4}, but for the $\Lambda$CDM model and
with $n_{\rm eff}$ ranging from 0 to 4 with a step of 1 (from the
bottom to the top).}
\label{cl_5}
\end{figure}
\begin{figure}
\begin{center}
\leavevmode
\begingroup%
  \makeatletter%
  \newcommand{\GNUPLOTspecial}{%
    \@sanitize\catcode`\%=14\relax\special}%
  \setlength{\unitlength}{0.065bp}%
\begin{picture}(3600,2000)(0,0)%
\special{psfile=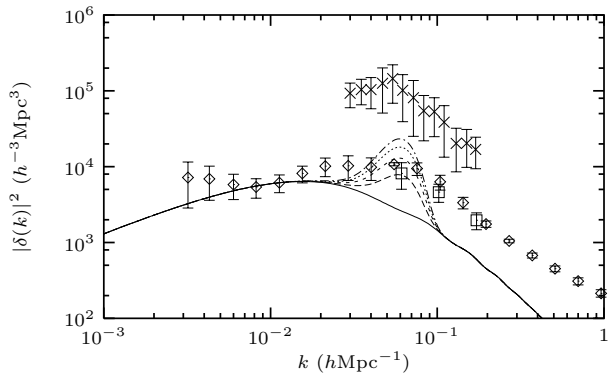 llx=0 lly=0 urx=720 ury=504 rwi=4680}
\put(2000,50){\makebox(0,0){\scriptsize  $k$ ($h$Mpc$^{-1}$)}}%
\put(100,1180){%
\special{ps: gsave currentpoint currentpoint translate
270 rotate neg exch neg exch translate}%
\makebox(0,0)[b]{\shortstack{\scriptsize $|\delta(k)|^2$ ($h^{-3}$Mpc$^3$)}}%
\special{ps: currentpoint grestore moveto}%
}%
\put(3450,200){\makebox(0,0){\scriptsize 1}}%
\put(2483,200){\makebox(0,0){\scriptsize $10^{-1}$}}%
\put(1517,200){\makebox(0,0){\scriptsize $10^{-2}$}}%
\put(550,200){\makebox(0,0){\scriptsize $10^{-3}$}}%
\put(500,2060){\makebox(0,0)[r]{\scriptsize $10^6$}}%
\put(500,1620){\makebox(0,0)[r]{\scriptsize $10^5$}}%
\put(500,1180){\makebox(0,0)[r]{\scriptsize $10^4$}}%
\put(500,740){\makebox(0,0)[r]{\scriptsize $10^3$}}%
\put(500,300){\makebox(0,0)[r]{\scriptsize $10^2$}}%
\end{picture}%
\endgroup
 
\end{center}
\caption{Same as Fig. \ref{ps_4}, but for the $\Lambda$CDM model and
with $n_{\rm eff}$ ranging from 0 to 4 with a step of 1 (from the
bottom to the top).}
\label{ps_5}
\end{figure}

Finally, in Figs.~\ref{cl_5} and \ref{ps_5} we show the CMBR anisotropies
and the matter power spectrum for the $\Lambda$CDM model including
both scalar and tensor contributions.
The same conclusions as for the SCDM model hold, but we note again
that, as for the case without gravitational waves, matter power
spectrum data favour a higher value of $n_{\rm eff}$ than CMBR
anisotropies data.
\par
We emphasize that when the gravitational waves contribution is not
negligible, the standard case $n_{\rm eff} = 0$ is excluded and that
extra power in the initial state is necessary.

\section{Conclusions}

In this paper we address the question of whether non-vacuum initial
states for cosmological perturbations are allowed, or whether they are
ruled out on the basis of present experimental and observational
data. 
\par 
The choice of the initial quantum state in which the quantum fields are
placed should be made on the basis of full quantum gravity. Since this
theory is at present unknown, we believe, as we discussed in the
Introduction, that it is worth studying non-vacuum initial states for
cosmological perturbations. Our choice of a non-vacuum initial state
is guided by the idea that the initial state could have a built-in
characteristic scale. We examined three 
different non-vacuum states,
which are compatible with the assumption of isotropy of the
Universe. Of particular interest is our choice of state $|\Psi_3
\rangle $, which seems to be the most natural rotational-invariant
smooth quantum state, which privileges a scale. We calculated the power
spectra of the Bardeen potential for these three states and compared
their theoretical predictions with current experimental and
observational data, namely the CMBR anisotropy measurements and the
redshift surveys of the distribution of galaxies.  With our choice of
initial states, the power spectra of the Bardeen potential possess a
peak, around the wave number, that corresponds to the built-in
characteristic scale of our model. The height of the peak is
controlled by the number of quanta $n$ of the initial state and its
width by another free parameter of our model. If the initial state 
is a quantum superposition then the height of the peak is 
controlled by the number $n_{\rm eff}$, which does not need 
to be an integer. 
\par 
The angular power spectrum of CMBR anisotropies for a model with
vanishing cosmological constant, tells us that the
characteristics of the first acoustic peak, as revealed by the
Saskatoon experiment, are compatible with the case $n_{\rm
eff}=1$. In the presence of a cosmological constant, CMBR anisotropy
measurements are in agreement with $n_{\rm eff}=0$ or $n_{\rm
eff}=1$, depending on the value of the cosmological parameters. The
observational data for the matter power spectra, as given by Einasto
{\it et al.}, favour higher values of the number of $n_{\rm eff}$ (2
or 3), whatever the value of the cosmological constant.
\par
The most realistic case is the one for which the sum of scalar and
tensor modes contributions is included. Considering standard
inflationary predictions for gravitational waves, we find that CMBR
anisotropies measurements require a higher value of $n_{\rm eff}$ (3
or 4) for both types of models, with and without a cosmological
constant, than in the case of an absence of tensor modes
contribution. This is in agreement with the matter power spectra. The
analysis of the redshift surveys by Einasto {\it et al.} leads
to matter power spectra that favour higher values of $n_{\rm eff}$,
once tensor contributions are also included. The interpretation of
these results for the states $|\Psi _2\rangle $ and $|\Psi _3\rangle $
leads to the conclusion that since 
 $n$ and  $n_{\rm eff}$ cannot be higher than a few, these states must be 
close to the vacuum. 
\par 
In conclusion, if the initial state of the cosmological perturbations is
not the vacuum but, instead, has a built-in characteristic scale, then
generic predictions of the model are: a high amplitude of the first
acoustic peak, a non-trivial feature in the matter power spectrum, and
deviations from Gaussianity in the CMBR map. It is too early to say
whether the results of the Saskatoon experiment (see also
Ref. \cite{PythonV}), as well as the analysis performed recently by
Einasto {\it et al.}, are first steps in this direction. More data are
needed and future experiments will be important in determining whether
the class of models proposed here provides an explanation which allows
a better description of the observations than the standard paradigm of
slow roll inflation plus cold dark matter.

\vspace{0.5cm}
\noindent

\acknowledgements
It is a pleasure to thank Robert Brandenberger for useful exchanges of
comments. Discussions with  Nathalie Deruelle, Ruth Durrer, Alejandro Gangui
and David Langlois are also acknowledged. 
We would also like to thank Volker M\"uller, who provided us with the cluster 
data.

\par{\tt martin@edelweiss.obspm.fr}
\par{\tt Alain.Riazuelo@obspm.fr}
\par{\tt Mairi.Sakellariadou@cern.ch}

\end{document}